\documentclass[12pt]{article}
\usepackage[dvips]{graphicx}
\usepackage{float}
\newcommand{\beq}{\begin{equation}}
\newcommand{\eeq}{\end{equation}}
\newcommand{\bdm}{\begin{displaymath}}
\newcommand{\edm}{\end{displaymath}}
\newcommand{\beqr}{\begin{eqnarray}}
\newcommand{\eeqr}{\end{eqnarray}}
\newcommand{\beqrn}{\begin{eqnarray*}}
\newcommand{\eeqrn}{\end{eqnarray*}}

\begin{document}
\title{Self-Dual Vortices in Abelian Higgs Models  with Dielectric Function on the Noncommutative Plane}
\author{W. Garc\'{\i}a Fuertes$^*$,\\
J. Mateos Guilarte$^\dag$}
\date{}
\maketitle
\begin{center}
$^*${{\it Departamento de F\'{\i}sica, Facultad de Ciencias,  Universidad de Oviedo,\\
E-33007 Oviedo, Spain.}}\\
$^\dag${{\it Departamento de F\'{\i}sica Fundamental and IUFFyM,  Universidad de Salamanca, 
E-37008 Salamanca, Spain.}}
\vskip1cm
\end{center}
\vskip1cm
\begin{abstract}
\noindent
We show that Abelian Higgs Models with dielectric function defined on the noncommutative plane enjoy self-dual vorticial solutions. By choosing a particular form of the dielectric function, we provide a family of solutions whose Higgs and magnetic fields interpolate between the profiles of the noncommutative Nielsen-Olesen and Chern-Simons vortices. This is done both for the usual $U(1)$ model and for the $SU(2)\times U(1)$ semilocal model with a doublet of complex scalar fields. The variety of known noncommutative self-dual vortices which display a regular behaviour when the noncommutativity parameter tends to zero results in this way considerably enlarged.
\end{abstract}
\vfill\eject
\section{Introduction}
Although local quantum field theory has had an impressive success as a framework for describing the dynamics of elementary particles at the current accesible energies, there are indications that, at some stage in the route towards a more fundamental theory, the idea of locality as a basic assumption of physics should  be given up. The exact way in which nonlocality would arise in that underlying theory is not clear, but a possibility that has often been considered by theorists is that, for lengths below some scale $\sqrt{\theta}$, spacetime has to be replaced by a different, blurred entity, in which the coordinates $x^\mu$ become noncommuting quantities $\hat{x}^\mu$ with commutators among then of order $\theta$. The reasons for considering noncommutative quantum field theories formulated on this arena are diverse. Originally, noncommutative QFT's appeared in an attempt to use the scale $\sqrt{\theta}$ as a cutoff for ultraviolet divergences, but later they were seen as effective theories on the spacetime foam resulting from the modified uncertainty principle arising in quantum gravity, as some low energy limits of the theory of open strings propagating on a constant Kalb-Ramond field or as describing the low energy quantum fluctuations of stacks of $D$-branes in the context of the IIB matrix model. The noncommutativity of spatial coordinates emerges also in condensed matter contexts, such as the motion of very light charged particles in strong magnetic fields as it happens in the quantum Hall effect. For reviews of the formalism of noncommutative quantum field theory and some of its motivations and uses or their possible role in phenomenology, see \cite {done}, \cite{szab}, \cite{jack}, \cite{vass}.

The study of the different classes of solitons appearing in field and string theory is an important topic, both because they are stable objects with interesting dynamical behaviour and because their conserved charges allow an interpretation of the solitons as supersymmetric BPS states, which can give significant information on the nonperturbative regime of the theory. In this respect, noncommutative QFT are especially appealing, because they can accomodate regular solitonic solutions in situations in which the usual commutative field theory would give singularities. This happens because Derrick's theorem, which is based on the scaling properties of the lagrangian kinetic and potential energy terms under dilatations of the coordinates, ceases to be valid in the noncommutative case due to the presence of the fundamental length $\sqrt{\theta}$. As a consequence, it is possible to find noncommutative scalar solitons even in theories without kinetic terms, and there is even a so-called {\it solution generating technique} which can be used to construct scalar and gauge solitons starting from trivial vacuum solutions \cite{harv}. This kind of solitons, however, become singular when the noncommutativity parameter $\theta$ is driven to zero.

In this paper we are going to study the self-dual vortices arising in a class of noncommutative Abelian models in which the kinetic Maxwell term incorporates a dielectric factor which is a function of the Higgs field. This dielectric contribution to the action, which spoils renormalizability, is however a common occurrence in the effective truncation to low-energy of supersymmetric theories. In the commutative case,  self-dual vortices in Abelian models with dielectric function have been studied in \cite{core}, \cite{baze}, \cite{ejc91} or \cite{ejc92}, and other related Higgs models which arise from effective supersymmetric theories are dealt with in \cite{susy97}, \cite{bhsm} and \cite{bchm}. Here, moving to the noncommutative plane, we will consider two variants among this kind of Abelian systems. First, we will pay attention to the case where there is only one complex scalar field and the local symmetry group is $U(1)$; this is the simplest paradigm for the Higgs mechanism and, from the phenomenological side,  has interest as a Ginzburg-Landau model for superconductivity (the scalar field is the order parameter between type I and II superconductivities). Then, we will extend the treatment to consider a model with a doublet of scalar fields enjoying a mixture of global $SU(2)$ and local $U(1)$ symmetries; this semilocal situation is a quite interesting limit of the electroweak theory and has been a subject of research in the field of cosmic strings. In both cases, we will focus on self-dual solutions wich continue to be regular when $\theta$ goes to zero. For that, we follow closely the treatment given in the articles \cite{lms1} and \cite{lms2} by Lozano, Moreno and Schaposnik. In these references, the authors solve the self-duality equations for, respectively, noncommutative Nielsen-Olesen and Chern-Simons-Higgs $U(1)$ vortices by means of a very convenient ansatz which leads to some discrete recurrence relations. On the other hand, in \cite{baze} a specific form of the dielectric function which interpolates between the commutative Nielsen-Olesen and Chern-Simons energy densities was proposed. We use this function (with a slightly different parametrization) to find the noncommutative vortices interpolating between those found in \cite{lms1} and \cite{lms2}, and also between their semilocal counterparts. The main theme of this paper is thus to combine the flexibility provided by a dielectric function with the techniques to deal with the noncommutative self-dual equations developed by the authors of \cite{lms1}, \cite{lms2} to show how the spectrum of self-dual noncommutative vortices with good behaviour for $\theta\rightarrow 0$ can be considerably enlarged.
\section{The Abelian Higgs model with dielectric function and its self-duality equations}
We are working on a three-dimensional spacetime with coordinates $(x^0,x^1,x^2)$ and metric $\eta_{\mu\nu}=\rm diag(1,-1,-1)$, but the spatial coordinates $x^1,x^2$ are not real numbers but fuzzy variables with uncertainty relation
\beq
\Delta x^1 \; \Delta x^2\geq \frac{\theta}{2} \label{eq:unc}
\eeq
where  $\theta$ is some positive real number. In this setup, we shall consider a dynamical model containing a complex scalar field $\phi$ and a gauge field $A_\mu$ interacting through the action
\bdm
S=\int d^3x \left\{-\frac{1}{4} G\ast F_{\mu\nu}\ast G\ast F^{\mu\nu}+D_\mu\phi \ast \overline{D^\mu\phi}-\frac{1}{2} W\ast W\right\} ,
\edm
where the star stands for the Groenewold-Moyal product
\beq
f(x)\ast g(x)=\exp\left[\frac{i}{2}\theta^{ij}\frac{\partial}{\partial x^i}\frac{\partial}{\partial x^{j\prime}} \right] \left. f(x) g(x^\prime)\right|_{x=x^\prime} \label{eq:moy}.
\eeq
The formalism of noncommutative gauge field theories is explained, for instance, in \cite{scha} or \cite{wagau}. In this particular model, the scalar field transforms with the fundamental representation of the $U_\ast (1)$ gauge group:
\bdm
\phi\longrightarrow \Lambda\ast\phi\hspace{2cm}\bar{\phi}\longrightarrow \bar\phi\ast\Lambda^\dagger ,
\edm
while $A_\mu$ is a $U_\ast (1)$ connection
\bdm
A_\mu\longrightarrow \Lambda\ast A_\mu\ast\Lambda^\dagger+\frac i e \Lambda\ast\partial_\mu \Lambda^\dagger ,
\edm
such that the scalar field covariant derivative and the gauge field strength are
\beqrn
D_\mu\phi&=&\partial_\mu \phi-i e A_\mu\ast\phi\\
F_{\mu\nu}&=&\partial_\mu A_\nu-\partial_\nu A_\mu-i e \left( A_\mu\ast A_\nu-A_\nu\ast A_\mu\right) .
\eeqrn
The field $\phi$ is self-interacting through a potential quadratic in $W$, a function of the star product of $\phi$ and $\bar\phi$, $W=W(\phi\ast\bar{\phi})$. Also, we allow for a non-minimal scalar-gauge interaction driven by the dielectric function $G=G(\phi\ast\bar{\phi})$. In this way,  $G$ and $W$ transforms under the adjoint representation of the gauge group
\bdm
G\longrightarrow \Lambda\ast G\ast \Lambda^\dagger\hspace{2cm}W\longrightarrow \Lambda\ast W\ast \Lambda^\dagger
\edm
exactly as $F_{\mu\nu}$ does, so that the gauge invariance of the action is guaranteed. In the following, we will also assume that $G$ is positive definite and that $W$ vanishes only when the product $\phi\ast\bar{\phi}$ takes its vacuum expectation value, denoted $v^2$.

Going to the temporal gauge $A_0=0$ and after some convenient rescalings
\bdm
A_\mu\rightarrow\frac{1}{e} A_\mu\hspace{1.5cm}\phi\rightarrow\frac{1}{e} \phi\hspace{1.5cm}v\rightarrow\frac{1}{e} v ,
\edm
the energy $E$ of the static field configurations takes the form
\bdm
e^2 E=\int d^2x \left\{\frac{1}{2} G\ast B\ast G\ast B+D_k\phi \ast \overline{D_k\phi}+\frac{1}{2} W\ast W\right\}
\edm
where $B$ is the magnetic field 
\bdm
B=F_{12}=\partial_1 A_2-\partial_2 A_2-i  \left( A_1\ast A_2-A_2\ast A_1\right)
\edm
and the spatial covariant derivatives are now $D_k \phi=\partial_k\phi-i A_k\ast\phi$, $k=1,2$. This form of the energy functional is amenable to a Bogomolny splitting. The quadratic term in the covariant derivatives of the Higgs field is written as \cite{jmwa}
\beq
\int d^2 x D_k\phi \ast \overline{D_k\phi}=\int d^2 x\left\{ \left( D_1 \phi+iD_2\phi\right)\ast \left( \overline{D_1 \phi}-i\overline{D_2\phi}\right)+\phi\ast\bar{\phi}\ast B\right\} \label{eq:bog1},
\eeq
where an irrelevant contour term has been discarded, and the other two terms can be combined as
\beq
\int d^2 x \left\{\frac{1}{2} G\ast B\ast G\ast B+\frac{1}{2}W\ast W\right\}=\int d^2 x \left\{\frac{1}{2} \left(G\ast B+W\right)^2-W\ast G\ast B\right\} \label{eq:bog2}, 
\eeq
where the square is in the sense of the $\ast$-operation and the ciclic property 
\bdm
\int d^2x\,f(x)\ast g(x)\ast\ h(x)=\int d^2x\, h(x)\ast f(x)\ast\ g(x)
\edm
of the Groenewold-Moyal product has been used. By combining expressions (\ref{eq:bog1}) and (\ref{eq:bog2}), we see that, if $W$ is chosen in such a way that
\beq
W\ast G=\phi\ast\bar{\phi}-v^2 ,\label{eq:cons}
\eeq
the energy of the field configurations which satisfy the self-duality equations
\beqr
G\ast B&=&-W \label{eq:dual1}\\
D_1 \phi+iD_2\phi&=&0\label{eq:dual2}
\eeqr
is proportional to the magnetic flux
\bdm
e^2 E=v^2 \int d^2 x B,
\edm
which is indeed a boundary term by virtue of
\bdm
\int d^2x\,A_1(x)\ast A_2(x)=\int d^2x\,A_2(x)\ast A_1(x).
\edm
For finite energy configurations, the fields at infinity depend only on the polar angle. The derivatives entering in (\ref{eq:moy}) are therefore proportional to inverse powers of distance and then, in the asymptotic region of the noncommutative plane, the star product of fields converges to the ordinary product. This means that the classification in topological sectors can be directly taken over from the well known results valid in the commutative plane. In particular, the magnetic flux is quantized. Hence, the solutions of (\ref{eq:dual1})-(\ref{eq:dual2}) minimize the energy in each topological sector
and are, therefore, bona fide solutions of the Euler-Lagrange equations. If we now denote as $\frac{1}{G}$ the inverse of $G$ according to the star product and take into account that $W$ and $G$ commute between themselves because both are functions of $\phi\ast\bar{\phi}$, the use of the constraint (\ref{eq:cons}) turns the first self-duality equation into the more convenient form 
\beq
B=\left(\frac{1}{G}\right)^2 \ast(v^2-\phi\ast\bar{\phi}) .
\eeq
We will use that form in what follows.

Functions on the noncommutative plane can be traded by operators on the Hilbert space ${\cal H}=L^2({\bf R^2})$ by means of the Weyl map
\bdm
f(x^1,x^2)\longrightarrow \hat{O}_f(\hat{x}^1,\hat{x}^2)=\frac{1}{(2 \pi)^2}\int d^2 k \hat{\Delta}(k) \tilde{f}(k),
\edm
where the Weyl kernel is $\hat\Delta (k)=\exp\left[-i (k_1\hat{x}_1+k_2\hat{x}_2)\right]$ and 

$
\tilde{f}(k)=\int d^2 x e^{i k\cdot x} f(x)
$
is the Fourier transform of $f(x)$, see \cite{harv, scha}. The transformation is consistent  in the sense that the star products are mapped to ordinary operator products on the Hilbert space. The use of the operator side of the Weyl map is very convenient for dealing with the self-duality equations, especially if we express them in holomorphic coordinates
\bdm
z=\frac{x^1+i x^2}{\sqrt{2}}\hspace{2cm}\bar{z}=\frac{x^1-i x^2}{\sqrt{2}}
\edm
and introduce the harmonic oscillator ladder operators
\bdm
\hat{a}=\frac{\hat{x}^1+i \hat{x}^2}{\sqrt{2\theta}}\hspace{2cm}\hat{a}^\dagger=\frac{\hat{x}^1+i \hat{x}^2}{\sqrt{2\theta}}
\edm
with commutator
\bdm
\left[\hat{a},\hat{a}^\dagger\right]=1 
\edm
consistent with the uncertainty relation (\ref{eq:unc}). One can check \cite{lms1,lms2} that, in terms of these operators, the self duality equations have the form
\beqrn
-\frac{1}{\sqrt{\theta}}\left[a^\dagger,A_{\bar{z}}\right]-\frac{1}{\sqrt{\theta}}\left[a,A_z\right]-i\left[A_z,A_{\bar{z}}\right]&=&\left(\frac{i}{G}\right)^2 (v^2-\phi\bar{\phi})\\
\frac{1}{\sqrt{\theta}}\left[a,\phi\right]-i A_{\bar{z}} \phi&=&0 
\eeqrn
whith $\phi$, $A_z$ and $A_{\bar{z}}$ representing here the operators $\hat{O}_\phi$, $\hat{O}_{A_z}$ and $\hat{O}_{A_{\bar{z}}}$ arising by applying the Weyl map to the Higgs and gauge fields of the original theory, but all hats have been supressed to alleviate notational cluttering. Also, the vortex energy can be now computed as the trace
\bdm
e^2 E=2\pi\theta v^2 \rm Tr_{\cal H} B
\edm 
on the Hilbert space.
\section{The interpolating model: noncommutative vortices}
By choosing the dielectric function in different forms it is possible to find self-dual noncommutative vortices with gauge and scalar fields displaying a wide variety of profiles. In particular, an interesting option proposed in \cite{baze} is to fix $G(\phi\ast\bar{\phi})$ in such a way that it can accommodate the profiles of the two most prominent types of vortices from a physical point of view: the Nielsen-Olesen and Chern-Simons vortices. This can be achieved by using
\bdm
G=\frac{1}{\sqrt{(1-\lambda)+\lambda\beta\phi\ast\bar{\phi}}}
\edm 
where the square root should be understood in the sense of the star product, $\lambda$ is a non-dimensional parameter with values in the interval $[0,1]$ and $\beta$ is an arbitrary constant with inverse mass squared dimension. Thus, the self dual equations for this model are
\beqrn
-\frac{1}{\sqrt{\theta}}\left[a^\dagger,A_{\bar{z}}\right]-\frac{1}{\sqrt{\theta}}\left[a,A_z\right]-i\left[A_z,A_{\bar{z}}\right]&=&i\left[(1-\lambda)+\lambda\beta\phi\bar{\phi}\right] (v^2-\phi\bar{\phi})\\
\frac{1}{\sqrt{\theta}}\left[a,\phi\right]-i A_{\bar{z}} \phi&=&0 .
\eeqrn
For $\lambda=0$, these equations are precisely the self-dual equations of the ordinary Abelian Higgs Model \cite{lms1}, while for $\lambda=1$ they coincide with those of the relativistic Chern-Simons-Higgs model \cite{lms2}, with the Chern-Simons $\kappa$ coupling given by $\kappa^2=\frac{1}{2\beta}$. Thus, by continously varying $\lambda$ between 0 and 1 we can find vortices with field-profiles which interpolate between the solutions arising in these two theories.
\subsection{Solving the noncommutative vortex equation}
Let us first consider, following \cite{jmwa} where more details can be found, the case of very large noncommutative parameter $\theta$. By expanding in inverse powers of $\theta$
\beqrn
\phi&=&\phi_\infty+\frac{1}{\theta}\phi_{-1}+\ldots\\
A_{\bar{z}}&=&\frac{1}{\sqrt{\theta}}\left((A_{\bar{z}})_\infty+\frac{1}{\theta}(A_{\bar{z}})_{-1}+\ldots\right)
\eeqrn
the self-dual equations are, to leading order, exactly the same that for Nielsen-Olesen vortices
\beqrn
\phi_\infty \bar{\phi}_\infty&=&v^2\\
i (A_{\bar{z}})_\infty&=&\left[a,\phi_\infty\right].
\eeqrn
As it is well known \cite{jmwa,witt}, these equations have a solution for each positive integer $n$ which can be expressed in terms of the shift operators $\mid k\rangle\langle k+n\mid$ for the harmonic oscillator:
\beqr
\phi_\infty&=&v\sum_{k=0}^\infty\mid k\rangle\langle k+n\mid\label{eq:inf1}\\
(A_{\bar{z}})_\infty&=&i\sum_{k=0}^\infty\left(\sqrt{k+1+n}-\sqrt{k+1}\right)\mid k\rangle\langle k+1\mid \label{eq:inf2}.
\eeqr
Because $a^n\mid k+n\rangle=\sqrt{(k+n)(k+n-1)\cdots (k+1)}\mid k\rangle$, the scalar field operator can be recast as
\bdm
\phi_\infty=\frac{v}{\sqrt{a^n (a^\dagger)^n}}a^n
\edm
and, in this way, the vorticial character of the solution is apparent through the factor $a^n$ (which is the noncommutative guise of the familiar angular dependence of type $z^n$ for commutative vortices). This character can be corroborated by computing the magnetic field, which is proportional to the projector onto the $\mid 0\rangle$ state,
\bdm
B_\infty=-i F_{z\bar{z}}=\frac{i}{\theta}\left[a^\dagger,(A_{\bar{z}})_\infty\right]+\frac{i}{\theta}\left[a,(A_z)_\infty\right]-\frac{1}{\theta}\left[(A_z)_\infty,(A_{\bar{z})_\infty}\right]=\frac{n}{\theta}\mid 0\rangle\langle 0\mid\ ,
\edm
and thus checking that the solution contains $n$ quanta of the magnetic flux
\bdm
\Phi_M=2\pi\theta {\rm Tr}_{\cal H} B_\infty=2\pi n
\edm
as it is appropriate for a vortex. 

However, the presence of $\theta$ in the denominator of the magnetic field shows that these solutions will become singular if we try to extend them to the commutative $\theta=0$ case. In order to obtain a solution valid for all values of $\theta$, it is natural to modify the solution (\ref{eq:inf1})-(\ref{eq:inf2}) for the $\theta=\infty$ case by trying an ansatz with a different coefficient for each shift operator
\beqr
\phi&=&v\sum_{k=0}^\infty f_k \mid k\rangle\langle k+n\mid\label{ans1}\\
A_{\bar{z}}&=&-\frac{i}{\sqrt{\theta}}\sum_{k=0}^\infty d_k \mid k\rangle\langle k+1\mid \label{ans2}
\eeqr
which was proposed for the Abelian Higgs Models in \cite{lms1} and for the Chern-Simons Higgs Model in \cite{lms2}.
By substitution in the self-dual equations, one finds a system of algebraic equations for the $f_k$ and $d_k$ coefficients 
\beqrn
d_k f_{k+1}&=&\sqrt{k+1}f_{k+1}-\sqrt{k+n+1}f_k\\
2\sqrt{k} d_{k-1}&-&2\sqrt{k+1} d_k+d_k^2-d^2_{k-1}=\theta v^2 (1-\lambda+\lambda\beta v^2 f_k^2)(1-f_k^2)
\eeqrn
which can be solved along the lines explained in these references. By writting $d_k$ as $d_k=\sqrt{k+1}-\sqrt{k+n+1}-e_k$ the first equation gives the new coefficient $e_k$ in terms of the $f_j$ coefficients as
\beq
e_k=\sqrt{k+n+1}\left(1-\frac{f_k}{f_{k+1}}\right) \label{eq:ek}
\eeq
and, with this expression for $e_k$, the second equation yields a three term recurrence relation for the $f_k$
\bdm
f_{k+1}^2 \left[(k+n) f_{k-1}^2+f_k^2(1+\theta v^2(1-\lambda+\lambda\beta v^2 f_k^2)(1-f_k^2))\right]=(k+n+1)f_k^4 .
\edm
which gives $f_{k+1}$ in terms of $f_k$ and $f_{k-1}$. As, on the other hand, $f_{-1}=0$, $f_1$ is only function of $f_0$
\bdm
f_1^2=\frac{(n+1) f_0^2}{1+\theta v^2(1-\lambda+\lambda\beta v^2 f_0^2)(1-f_0^2)} .
\edm
Thus, once $f_0$ is chosen and $f_1$ determined, all the remaining coefficients can be recursively found. The task is to find the value of $f_0$ which matches the boundary condition for $k\rightarrow\infty$: in this limit, the $f_k$ have to approach unity, which is the only fixed point of the recurrence relation, and to accomplish it a simple bisection method can be used:  we try first with $f_0^2=0.5$; if we find that $f_k$ grows over unity before $f_k<f_{k-1}$, $f_0$ is too large and we change $f_0$ to $f_0^2=0.25$; if, instead, $f_k<f_{k-1}$ before $f_k>1$, $f_0$ is too small and we try with $f_0^2=0.75$. We repeat this procedure until a good matching with the boundary condition is attained. Once $f_0$ and all the the $f_k$ coefficients are known, the magnetic field can be calculated from the self-duality equations as
\bdm
B=v^2\sum_{k=0}^\infty \left[1- \lambda+\lambda \beta v^2 f_k^2\right](1-f_k^2) \mid k\rangle\langle k \mid
\edm
and, thus, the magnetic flux and energy are
\beq
\Phi_M=2\pi\theta {\rm Tr}_{\cal H} B=2 \pi\theta v^2 \sum_{k=0}^\infty \left[1- \lambda+\lambda \beta v^2 f_k^2\right](1-f_k^2) \label{eq:mag}
\eeq
and
\bdm
E=\frac{v^2}{e^2} \Phi_M .
\edm
In fact, for topological reasons, we expect that $\Phi_M=2\pi n$, irrespective of the values of $\theta, \lambda, \beta$ or $v$, for any solution of the self-duality equations.

We have computed the correct value of $f_0$ for several values of the parameters in the dielectric function and for topological number $n=1$. For the case $\lambda=0$, $\beta$ disappears form the action and only the non-dimensional combination $\theta v^2$ matters. The following table shows the results
{\scriptsize
\begin{center}
\begin{tabular}{||c|c|c|c|c|c|c|c|c||}
\hline
\multicolumn{9}{||c||}{$\lambda=0$}\\
\hline
$\theta v^2$&0.25&0.50&0.75&1.00&1.25&1.50&1.75&2.00\\
\hline
$f_0^2$&0.2572165&0.4006888&0.4940118&0.5602955&0.6101472&0.6491837&0.6806831&0.7066985 \\
\hline
\end{tabular}
\end{center}}
\noindent For the other extreme case $\lambda=1$, the parameters merge in a global factor $\beta\theta v^4$. The results are
{\scriptsize
\begin{center}
\begin{tabular}{||c|c|c|c|c|c|c|c|c||}
\hline
\multicolumn{9}{||c||}{$\lambda=1$}\\
\hline
$\beta\theta v^4$&0.25&0.50&0.75&1.00&1.25&1.50&1.75&2.00\\
\hline
$f_0^2$&0.1082514  &0.2168143  &0.3170487  &0.4037747  &0.4758857  &0.5348773  &0.5831093  &0.6228436 \\
\hline
\end{tabular}
\end{center}}
\noindent The numbers in the two previous tables are in good agreement with those appearing in \cite{lms1}, \cite{lms2}. In the next four ones we present the results for $f_0^2$ for some values of $\lambda$ interpolating between the Nielsen-Olesen and Chern-Simons cases. In the tables, the rows and columns correspond, respectively, to the values of $\beta v^2$ and $\theta v^2$ given in the margins.
{\scriptsize
\begin{center}
\begin{tabular}{||c|c|c|c|c|c|c|c|c||}
\hline
\multicolumn{9}{||c||}{$\lambda=0.2$}\\
\hline
$\beta v^2 \downarrow/ \theta v^2\rightarrow$&0.25&0.50&0.75&1.00&1.25&1.50&1.75&2.00\\
\hline
 0.25   &0.2225542   &0.3578602   &0.4500099   &0.5173522   &0.5689932   &0.6100056   &0.6434570   &0.6713201\\
 \hline
  0.50   &0.2267165   &0.3642712   &0.4576113   &0.5255624   &0.5774821   &0.6185802   &0.6520019   &0.6797658\\
  \hline
  0.75   &0.2308680   &0.3706328   &0.4651091   &0.5336139   &0.5857629   &0.6269045   &0.6602619   &0.6878986\\
  \hline
 1.00   &0.2350088   &0.3769440   &0.4725019   &0.5415063   &0.5938368   &0.6349824   &0.6682435   &0.6957276\\
 \hline
 1.25   &0.2391388   &0.3832038   &0.4797885   &0.5492392   &0.6017055   &0.6428179   &0.6759534   &0.7032625\\
 \hline
 1.50   &0.2432580   &0.3894111   &0.4869678   &0.5568126   &0.6093711   &0.6504157   &0.6833991   &0.7105131\\
 \hline
 1.75   &0.2473662   &0.3955651   &0.4940389   &0.5642271   &0.6168360   &0.6577808   &0.6905881   &0.7174894\\
\hline
 2.00   &0.2514633   &0.4016646   &0.5010010   &0.5714831   &0.6241032   &0.6649185   &0.6975281   &0.7242017\\
\hline
\end{tabular}
\end{center}}
{\scriptsize
\begin{center}
\begin{tabular}{||c|c|c|c|c|c|c|c|c||}
\hline
\multicolumn{9}{||c||}{$\lambda=0.4$}\\
\hline
$\beta v^2 \downarrow/ \theta v^2\rightarrow$&0.25&0.50&0.75&1.00&1.25&1.50&1.75&2.00\\
\hline
  0.25   &0.1834747   &0.3064902   &0.3952554   &0.4626214   &0.5156652   &0.5586206   &0.5941838   &0.6241573\\ \hline
  0.50   &0.1923526   &0.3208740   &0.4129240   &0.4821936   &0.5362822   &0.5797419   &0.6154649   &0.6453767\\ \hline
  0.75   &0.2011923   &0.3350540   &0.4301225   &0.5010010   &0.5558540   &0.5995703   &0.6352428   &0.6649185\\ \hline
 1.00   &0.2099940   &0.3490185   &0.4468313   &0.5190293   &0.5743853   &0.6181386   &0.6535835   &0.6828842\\ \hline
 1.25   &0.2187571   &0.3627553   &0.4630337   &0.5362727   &0.5918931   &0.6354940   &0.6705672   &0.6993862\\ \hline
 1.50   &0.2274800   &0.3762523   &0.4787170   &0.5527332   &0.6084050   &0.6516942   &0.6862819   &0.7145411\\ \hline
 1.75   &0.2361604   &0.3894982   &0.4938726   &0.5684202   &0.6239571   &0.6668039   &0.7008194   &0.7284646\\ \hline
 2.00   &0.2447957   &0.4024825   &0.5084961   &0.5833496   &0.6385915   &0.6808915   &0.7142714   &0.7412679\\ \hline
\end{tabular}
\end{center}}
{\scriptsize
\begin{center}
\begin{tabular}{||c|c|c|c|c|c|c|c|c||}
\hline
\multicolumn{9}{||c||}{$\lambda=0.6$}\\
\hline
$\beta v^2 \downarrow/ \theta v^2\rightarrow$&0.25&0.50&0.75&1.00&1.25&1.50&1.75&2.00\\
\hline
  0.25   &0.1389372   &0.2432580   &0.3244385   &0.3894111   &0.4426027   &0.4869678   &0.5245515   &0.5568126\\ \hline
  0.50   &0.1531621   &0.2677365   &0.3559161   &0.4255004   &0.4816218   &0.5277492   &0.5662879   &0.5989462\\ \hline
  0.75   &0.1673229   &0.2917726   &0.3862109   &0.4594802   &0.5175747   &0.5645702   &0.6032743   &0.6356563\\ \hline
 1.00   &0.1814214   &0.3153038   &0.4151959   &0.4912258   &0.5504218   &0.5975425   &0.6358143   &0.6674566\\ \hline
 1.25   &0.1954515   &0.3382635   &0.4427719   &0.5206897   &0.5802441   &0.6269188   &0.6643456   &0.6949639\\ \hline
 1.50   &0.2094023   &0.3605889   &0.4688750   &0.5478960   &0.6072109   &0.6530306   &0.6893532   &0.7187975\\ \hline
 1.75   &0.2232606   &0.3822249   &0.4934771   &0.5729255   &0.6315447   &0.6762369   &0.7113119   &0.7395235\\ \hline
 2.00   &0.2370120   &0.4031266   &0.5165834   &0.5958987   &0.6534912   &0.6968901   &0.7306554   &0.7576340\\ \hline
\end{tabular}
\end{center}}
{\scriptsize
\begin{center}
\begin{tabular}{||c|c|c|c|c|c|c|c|c||}
\hline
\multicolumn{9}{||c||}{$\lambda=0.8$}\\
\hline
$\beta v^2 \downarrow/ \theta v^2\rightarrow$&0.25&0.50&0.75&1.00&1.25&1.50&1.75&2.00\\
\hline
  0.25   &0.0874802   &0.1626094   &0.2274800   &0.2838134   &0.3330198   &0.3762523   &0.4144539   &0.4483972\\ \hline
  0.50   &0.1077297   &0.2001111   &0.2788254   &0.3457787   &0.4028276   &0.4516387   &0.4936371   &0.5300064\\ \hline
  0.75   &0.1279574   &0.2370120   &0.3279805   &0.4031266   &0.4651461   &0.5165834   &0.5595923   &0.5958986\\ \hline
 1.00   &0.1481567   &0.2730445   &0.3743037   &0.4550554   &0.5194259   &0.5711959   &0.6133762   &0.6482294\\ \hline
 1.25   &0.1682918   &0.3079312   &0.4173346   &0.5013053   &0.5660076   &0.6166414   &0.6570262   &0.6898471\\ \hline
 1.50   &0.1883157   &0.3414295   &0.4568418   &0.5420656   &0.6057372   &0.6544386   &0.6926365   &0.7232946\\ \hline
 1.75   &0.2081774   &0.3733515   &0.4928022   &0.5778005   &0.6396228   &0.6860415   &0.7219792   &0.7505531\\ \hline
 2.00   &0.2278258   &0.4035707   &0.5253498   &0.6090921   &0.6686338   &0.7126815   &0.7464413   &0.7730916\\ \hline
\end{tabular}
\end{center}}
\subsection{Comparison with the commutative vortices}
For all cases shown in the previous subsection, we have checked using (\ref{eq:mag}) that the magnetic flux takes the value $\Phi_M=2\pi $,  as it should be. As was done in \cite{lms1}, it is also interesting to check if the vortices of the noncommutative model converge to those of the commutative one when the parameter $\theta$ goes to zero. Using the ansatz
\beqrn
\phi&=&v g(r) e^{i n \varphi}\\
A_\theta&=&n-\alpha(r)
\eeqrn
with $r$ and $\varphi$ the standard polar coordinates, the self-duality equations of the commutative model are \cite{core,baze,ejc91}
\beqrn
\frac{1}{r}\frac{d\alpha}{dr}&=&\left(1-\lambda+\lambda\beta v^2 g^2\right)(g^2-1)\\
\frac{dg}{dr}&=&\frac{\alpha g}{r} ,
\eeqrn 
and the boundary conditions are 
\beqrn
g(0)=0&\hspace{3cm}&g(\infty)=1\\
\alpha(0)=n&\hspace{3cm}&\alpha(\infty)=0 .
\eeqrn
For $r\simeq 0$, the solution is
\beqrn
g(r)&\simeq&g_0 r^n\\
\alpha(r)&\simeq&n+\frac{\lambda-1}{2} r^2
\eeqrn
and starting with this asymptotics, the equations can be integrated numerically to find the value of $g_0$ which matches the boundary conditions at infinity. We have done this with a fourth-order Runge-Kutta method for different values of $\lambda$ and $\beta v^2$ and found the following results for $g_0^2$:
{\scriptsize
\begin{center}
\begin{tabular}{||c|c|c|c|c|c|c|c|c||}
\hline
$\lambda\downarrow / \beta v^2\rightarrow$&0.25&0.50&0.75&1.00&1.25&1.50&1.75&2.00\\
\hline
0.0&0.7279&0.7279&0.7279&0.7279&0.7279&0.7279&0.7279&0.7279 \\
\hline
0.2&0.5933&0.6042&0.6152&0.6260&0.6369&0.6478&0.6586&0.6695 \\
\hline
0.4&0.4586&0.4804&0.5021&0.5237&0.5453&0.5668&0.5882&0.6096 \\
\hline
0.6&0.3239&0.3563&0.3886&0.4207&0.4527&0.4846&0.5164&0.5482 \\
\hline
0.8&0.1889&0.2317&0.2741&0.3164&0.3586&0.4008&0.4429&0.4850 \\
\hline
1.0&0.0524&0.1049&0.1573&0.2098&0.2622&0.3147&0.3671&0.4195 \\
\hline
\end{tabular}
\end{center}}
We have, on the other hand, computed $f_0^2$ for very small $\theta$ for the diverse values of $\beta v^2$ and $\lambda$ shown in the previous table and we find a perfect agreement between $g_0^2$ in the commutative model and $\frac{f_0^2}{2\theta v^2}$ in the noncommutative one, exactly as was established for the case $G=1$ in \cite{lms1}. To understand this coincidence, let us write the scalar field of the noncommutative $n=1$ vortex as
\bdm
\phi=\frac{v}{\sqrt{a a^\dagger}}f(a^\dagger a) a\hspace{1cm}{\rm with}\hspace{1cm}f(a^\dagger a)\mid k\rangle=f_k\mid k\rangle
\edm
and compute his expected value on the coherent state $\mid w\rangle =e^{-\frac{\mid w\mid^2}{2}} e^{w a^\dagger}\mid 0\rangle$ which satisfies $\langle w\mid x^1+i x^2 \mid w \rangle=\sqrt{2\theta}\left( {\rm Re}\, w+ i\, {\rm Im}\, w \right)$ and represents a minimal wavepacket centered around the point $\sqrt{2\theta}\, w$ of the noncommutative plane with spread $\Delta x^1=\Delta x^2=\sqrt{\frac{\theta}{2}}$ \cite{jmwa}. Now, using
\bdm
\langle w \mid\phi\mid w \rangle =\langle 0\mid \frac{v}{\sqrt{(a+w) (a^\dagger+\bar{w})}}f((a^\dagger+\bar{w}) (a+w))(a+w)\mid 0\rangle
\edm
we see that for $w\rightarrow 0$ and in the limit $\theta\rightarrow 0$ we have 
\bdm
\langle w \mid\phi\mid w\rangle\rightarrow v f_0 w=\frac{v f_0}{\sqrt{2 \theta}}(x^1+i x^2)
\edm
and this should be interpreted as the value of $\phi$ near the origin. On the other hand, for the commutative model $\phi\simeq g_0 v^2 r e^\varphi$, so one should expect $g_0^2=\frac{f_0^2}{2\theta v^2}$ as it indeed occurs.
\subsection{Noncommutative vortex profiles}
Once the scalar and magnetic field operators are known in Hilbert space, it is not difficult to invert the Weyl map and find the functional form of these vortex fields in the noncommutative coordinates. For that, we only have to take into account that the function $f_{j,k}(x)$ whith Weyl transform $\mid j\rangle\langle k \mid$ is \cite{harv}
\bdm
f_{j,k}(x)=2 (-1)^j \sqrt{\frac{j!}{k!}}e^{-\frac{r^2}{\theta}}\left(2\frac{r^2}{\theta}\right)^\frac{j-k}{k} L_{j}^{k-j}\left(2\frac{r^2}{\theta}\right)e^{i(k-j)\varphi}
\edm
where $z=\frac{r}{\sqrt{2}}\exp{i\varphi}$ and the $L_p^q(y)$ are generalized Laguerre polynomials. In particular, as
\bdm
\phi\bar{\phi}=v^2\sum_{k=0}^\infty f_k^2\mid k\rangle\langle k\mid
\edm 
and the magnetic field is (\ref{eq:mag}) we find
\bdm
\phi(x)\ast\bar{\phi}(x)=v^2\sum_{k=0}^\infty f_k^2 f_{j,j}(x)
\edm
and
\bdm
B(x)=v^2\sum_{k=0}^\infty (1-\lambda+\lambda\beta v^2 f_k^2)(1-f_k^2) f_{j,j}(x)
\edm
The following figures show the profiles of $\phi\ast\bar{\phi}$ (in red) and $B$ (in green) as a function of $r$ for several values of the non-dimensional parameters $\theta v^2$ and $\beta v^2$. The curves in each figure are for $\lambda=0,\,0.2,\,0.4,\,0.6,\,0.8$ and 1, and one can distinguish these values because in all cases both $\phi\ast\bar\phi$ and $B$ at the origin decrease with $\lambda$.
\begin{figure}[H]
\centering
\includegraphics[width=7.5cm,height=7.5cm]{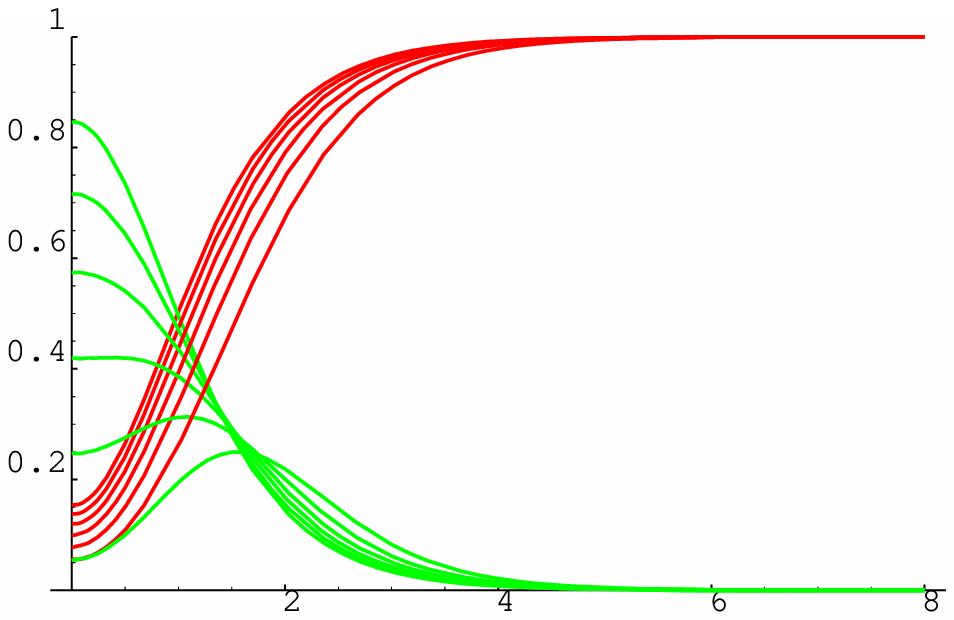}
\includegraphics[width=7.5cm,height=7.5cm]{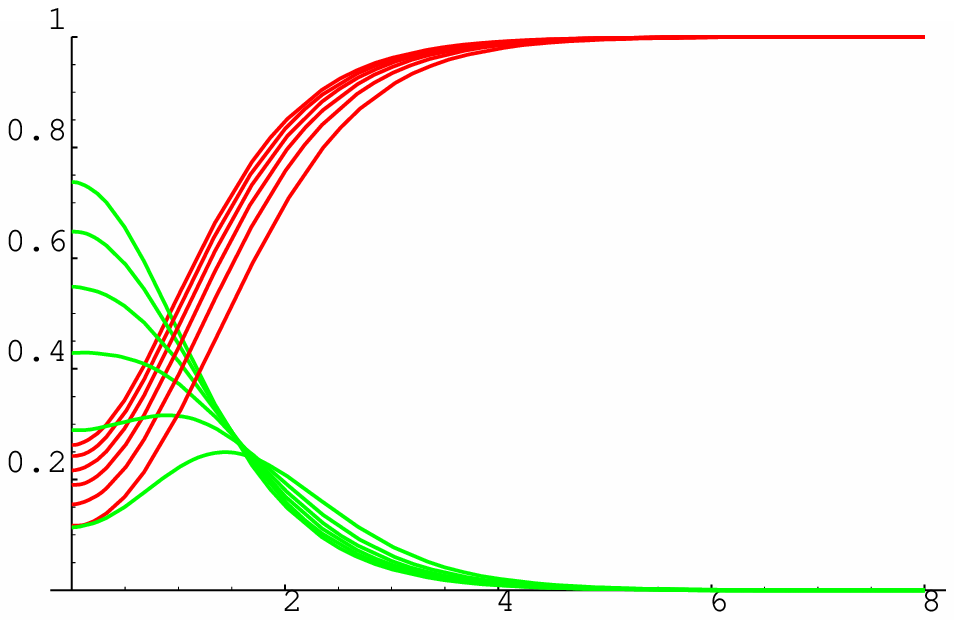}
\caption{The cases $\theta v^2=0.25,\beta v^2=1$ (left) and $\theta v^2=0.5,\beta v^2=1$ (right).}
\end{figure}

\begin{figure}[H]
\centering
\includegraphics[width=7.5cm,height=7.5cm]{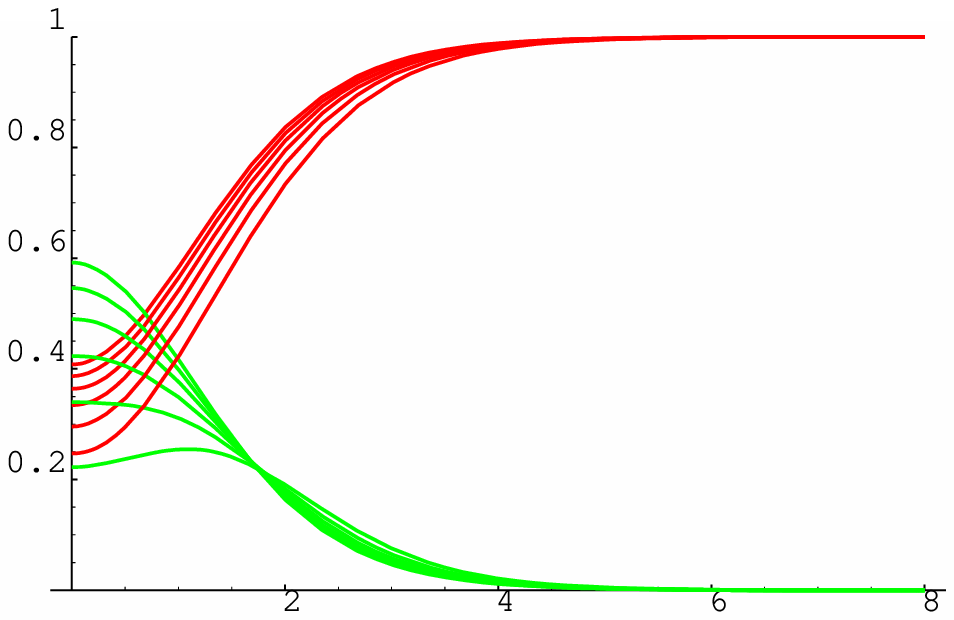}
\includegraphics[width=7.5cm,height=7.5cm]{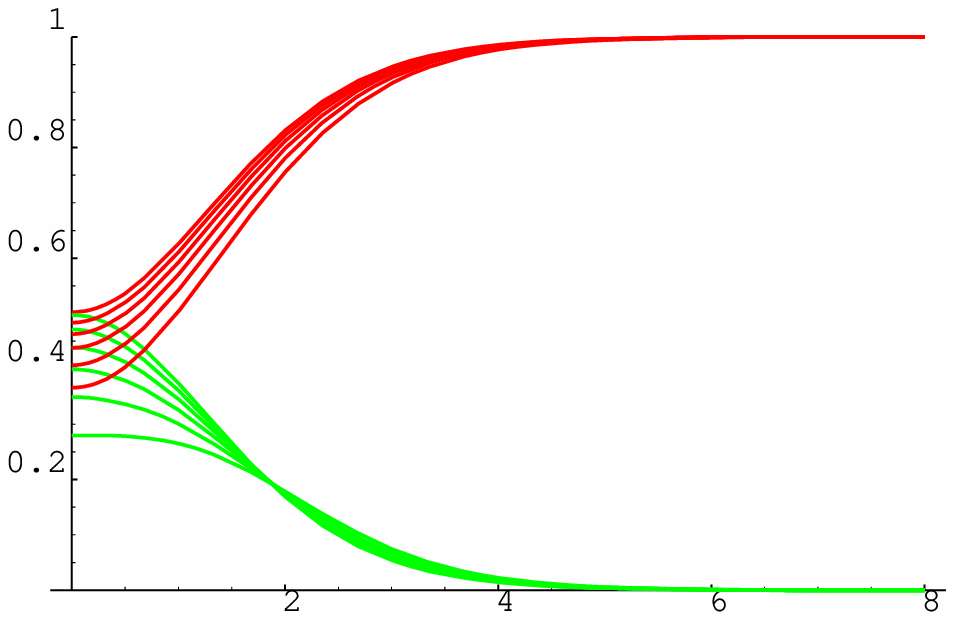}
\caption{The cases $\theta v^2=1,\beta v^2=1$ (left) and $\theta v^2=1.5,\beta v^2=1$ (right).}
\end{figure}
\begin{figure}[H]
\centering
\includegraphics[width=7.5cm,height=7.5cm]{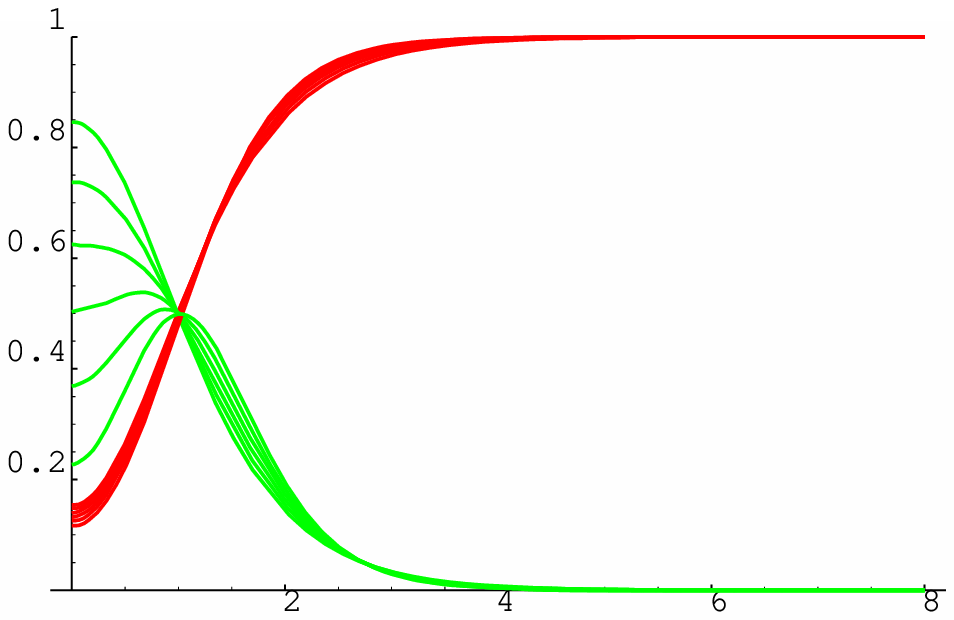}
\includegraphics[width=7.5cm,height=7.5cm]{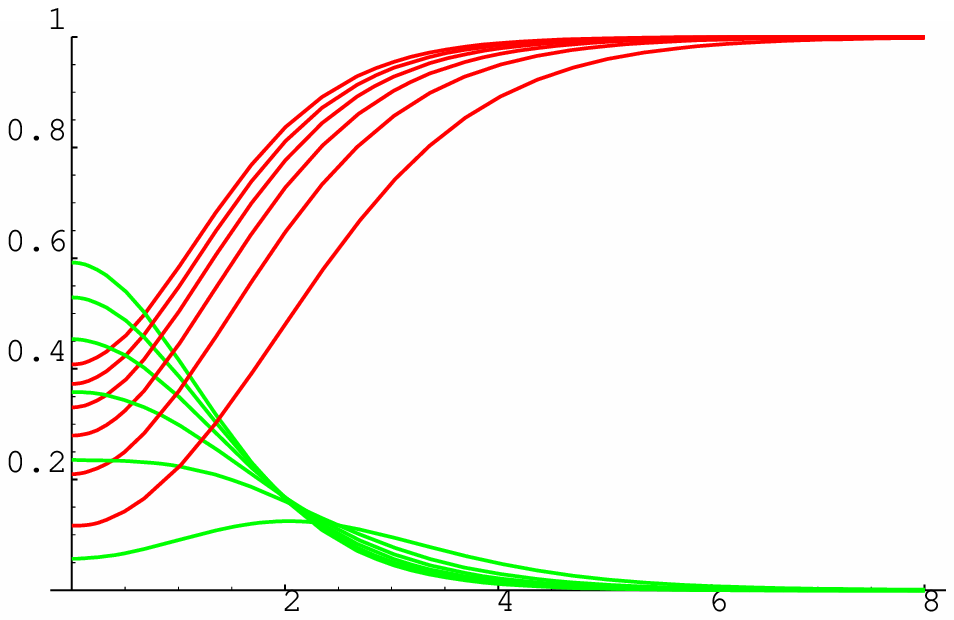}
\caption{The cases $\theta v^2=0.25,\beta v^2=2$ (left) and $\theta v^2=1,\beta v^2=0.5$ (right).}
\end{figure}
A few comments about these figures:
\begin{itemize}
\item As one can see, the profile of the magnetic field exhibits a maximum at the center of the Nielsen-Olesen vortex, which is more and more flat as $\lambda$ increases, and becomes finally a minimum for the Chern-Simons case. Thus, the magnetic field concentrates at a peak for small $\lambda$ and is more disperse, forming a ring around the core of the vortex, as $\lambda$ approaches one. The effect is more noticeable when the parameter $\theta$ measuring the noncommutativity of the plane is small; in particular, in the fourth figure, which has the larger value of $\theta v^2$, the magnetic field at the center of $\lambda=1$ case looks more like a plateu than like a ring. 
\item The first four figures have $\beta v^2$ fixed and a increasing degree of noncommutativity, with parameter varying from $\theta v^2=0.25$ to $\theta v^2=1.5$. Looking at these, we see that, as $\theta v^2$ increases, the value of $B(0)$ decreases for the Nielsen-Olesen vortices, but increases for the Chern-Simons ones. Thus, while for small noncommutativity the magnetic field at the center of the vortex shows a wide variation with $\lambda$, for higher $\theta v^2$ the range of this variation is of lesser extent. 
\item In the same four figures, we can see that the profiles of $\phi\ast\bar\phi$ with distance are quite similar for all the values of $\lambda$, but $\phi(0)\ast\bar\phi(0)$ increases with $\theta v^2$. This is as it should be expected, given that for commutative vortices the scalar field vanishes at the origin. 
\item The fifth figure is for small noncommutativity and large $\beta v^2$, which corresponds to small Chern-Simons parameter $\kappa$. There is in this case a very important variation of the magnetic field with $\lambda$, and the ring-like shape of the core of the vortex is evident for quite low values of the intepolating parameter. Instead, the dependence of the profile of $\phi\ast\bar\phi$ with $\lambda$ is completely negligible. 
\item This behavior is in contrast with the sixth figure, which has more amount of noncommutativity and a small $\beta v^2$. In this case, both the magnetic field $B$ and the scalar field magnitude $\phi\ast\bar\phi$ show substantial variations as we interpolate between the Nielsen-Olesen and Chern-Simons solutions.
\end{itemize}
\section{The semilocal model with dielectric function}
Unlike the AHM, the Standard Model of particle physics does not admit topologically stable vortices. The reason is that, in this case, the pattern of gauge symmetry breaking is $G=SU(2)\times U(1)\rightarrow H=U(1)$ and the fundamental group of the quotient $G/H$ is trivial, $\pi_1(G/H)=1$. There is, however, an interesting exception to this general statement: if the Weinberg angle is $\theta_W=\frac{\pi}{2}$, the weak isospin gauge bosons decouple, the $SU(2)$ factor becomes a global symmetry and stable flux lines appear in the spectrum \cite{vach}. Being the consequence of the mixing of the global $SU(2)$ and gauge $U(1)$ symmetries, these solutions are  kown as semilocal vortices. Although the Higgs field is a $SU(2)$ doublet
\bdm
\Phi=\left(\begin{array}{c}\phi^+\\\phi^0\end{array}\right)
\edm
and thus the vacuum orbit is $S^3$, the stability of semilocal vortices is guaranteed because, to ensure the vanishing of their covariant derivatives at long distances, the asymptotic scalar field has to be given by a map from the spatial $S^1$ border to one $S^1$ fiber of the Hopf fibration $S^3\rightarrow S^2$ \cite{gibb}, \cite{pos9}. Hence, the effective fundamental group which classifies the finite energy configurations is $\pi_1(S^1)={\bf Z}$, the winding number corresponding, as usual, to the magnetic flux. In each topological sector, the axially symmetric semilocal vortices form a family which is parametrized by a complex number and interpolates between standard Nielsen-Olesen vortices and ${\bf C}P^1$-lumps. Although all the defects in the family are stable \cite{hind}, the fields decay exponentially at infinity only for the Nielsen-Olesen vortices. For the other cases the magnetic flux is more spread and the fields reach their asymptotic values as inverse powers of the distance. 
\subsection{Semilocal self-dual noncommutative vortices}
Our aim in this section is to study the self-dual vortex solutions arising in a noncommutative semilocal model with dielectric function. With the rescalings seen in Sect.2, the action of the model is
\bdm
S=\int d^3x \left\{-\frac{1}{4} G\ast F_{\mu\nu}\ast G\ast F^{\mu\nu}+\sum_{a=+,0}D_\mu\phi^a \ast \overline{D^\mu\phi^a}-\frac{1}{2} W\ast W\right\} .
\edm
where $G$ and $W$ are positive functions which, to be covariant under both the global $SU(2)$ and the gauge $U(1)$ symmetries, have the structure:
\bdm
G=G(\sum_{a=0,+}\phi^a \bar{\phi}^a),\hspace{2cm}W=W(\sum_{a=0,+}\phi^a \bar{\phi}^a) .
\edm
The energy for static configurations is
\bdm
e^2 E=\int d^2x \left\{\frac{1}{2} G\ast B\ast G\ast B+\sum_{a=0,+}D_k\phi^a \ast \overline{D_k\phi^a}+\frac{1}{2} W\ast W\right\}
\edm
and, as in the previous section, one can perform a Bogomolny spliting such as, if
\beq
W\ast G=\sum_{a=+,0}\phi^a\ast\bar{\phi}^a-v^2\ , 
\eeq
the field configurations which satisfy the self-duality equations
\beqrn
G\ast B&=&-W\\
D_1 \phi^++iD_2\phi^+&=&0\\
D_1 \phi^0+iD_2\phi^0&=&0
\eeqrn
saturate the Bogomolny bound
\bdm
e^2 E\geq v^2 \int d^2 x B. 
\edm
As before, we choose a dielectric function
\bdm
G=\frac{1}{\sqrt{(1-\lambda)+\lambda\beta\sum_{a=+,0}\phi^a\ast\bar{\phi}^a}}
\edm 
which is suitable to interpolate between semilocal vortices of Maxwell type and semilocal Chern-Simons vortices; for the latter, see \cite{juwi} \footnote{Here we abide by the notation of \cite{lms2}. To compare with \cite{juwi}, $A_\mu,\partial_\mu, \phi,\kappa$ and $\eta$ in that paper have to be rescaled according to $A_\mu\rightarrow \frac{A_\mu}{\eta},x_\mu\rightarrow \eta x_\mu,\phi\rightarrow \frac{\sqrt{2} \phi}{\eta},\kappa\rightarrow \frac{2\kappa}{\eta},\eta\rightarrow \sqrt{2}\eta$.}. Then, the Bogomolny equations are
\beqr
-\frac{1}{\sqrt{\theta}}\left[a^\dagger,A_{\bar{z}}\right]-\frac{1}{\sqrt{\theta}}\left[a,A_z\right]-i\left[A_z,A_{\bar{z}}\right]&=&i\left[(1-\lambda)+\lambda\beta\sum_{a=+,0}\phi^a\bar{\phi}^a\right] (v^2-\sum_{a=+,0}\phi^a\bar{\phi}^a)\nonumber\\
\frac{1}{\sqrt{\theta}}\left[a,\phi^+\right]-i A_{\bar{z}} \phi^+&=&0\label{eq:mas}\\ 
\frac{1}{\sqrt{\theta}}\left[a,\phi^0\right]-i A_{\bar{z}} \phi^0&=&0\label{eq:cero} .
\eeqr
and a convenient ansatz to solve them in the sector of magnetic flux $\Phi_M=2\pi n$ is a direct extension of (\ref{ans1})-(\ref{ans2}):
\beqr
\phi^+&=&v\sum_{k=0}^\infty f_k \mid k\rangle\langle k+n\mid\\
\phi^0&=&v\sum_{k=0}^\infty \eta_k \mid k\rangle\langle k+l\mid\\
A_{\bar{z}}&=&-\frac{i}{\sqrt{\theta}}\sum_{k=0}^\infty d_k \mid k\rangle\langle k+1\mid\  ,
\eeqr
in which we have used the global $SU(2)$ symmetry to put the topological vorticity in the $\phi^+$ component, i.e. we will use boundary conditions $f_k\rightarrow 1$, $\eta_k\rightarrow 0$ for $k\rightarrow \infty$, but we also allow for a behavior of type $a^l$ for the other component. This mimics the angular dependence of the solutions found for the commutative model, see \cite{gibb}, \cite{hind}, and the analysis of that case suggests that, in order to have well-behaved finite energy solutions, we have to take $0\leq l\leq n-1$.  

Using this ansatz in (\ref{eq:mas}) and (\ref{eq:cero}), we can relate the coefficients $\eta_k$ and $f_k$ through
\bdm
\eta_{k+1}=\sqrt{\frac{k+l+1}{k+n+1}}\frac{f_{k+1}}{f_k} \eta_k
\edm
and then, iterating this relation and using the same arguments of the previous section, we see that, once some initial values for $f_0$ and $\eta_0$ are given, all remaining coefficients follow from the recurrence relations
\beqrn
\eta_k&=&\sqrt{\frac{(k+l)!\; n!}{(k+n)!\; l!}} \frac{f_k}{f_0} \eta_0\\ 
(k+n+1)f_k^4&=&\left[(k+n) f_{k-1}^2+f_k^2(1+\theta v^2(1-\lambda+\lambda\beta v^2 (f_k^2+\eta_k^2))(1-f_k^2-\eta_k^2))\right] f_{k+1}^2\\
\eeqrn
and equation (\ref{eq:ek}). Thus, the problem is to find, for each $\eta_0$, the value of $f_0^2$ which gives the correct behaviour for $k\rightarrow\infty$. Using the bisection method, we have found $f_0^2$ for $n=1, l=0$ and the cases given in the following tables:
{\scriptsize
\begin{center}
\begin{tabular}{||c|c|c|c|c|c|c|c|c||}
\hline
\multicolumn{9}{||c||}{$\lambda=0$}\\
\hline
$\eta_0\downarrow/ \theta v^2\rightarrow$&0.25&0.50&0.75&1.00&1.25&1.50&1.75&2.00\\
\hline
0.1&0.2542716&0.3962479&0.4886493&0.5543005&0.6036871&0.6423652&0.6735785&0.6993595\\ \hline
0.2&0.2454721&0.3829659&0.4726012&0.5363519&0.5843404&0.6219402&0.6522926&0.6773681\\ \hline
0.3&0.2309262&0.3609669&0.4459876&0.5065611&0.5522091&0.5880020&0.6169109&0.6408026\\ \hline
0.4&0.2108201&0.3304652&0.4090162&0.4651210&0.5074699&0.5407120&0.5675811&0.5897986\\ \hline
0.5&0.1854286&0.2917786&0.3619958&0.4123184&0.4503856&0.4803106&0.5045229&0.5245576\\ \hline
0.6&0.1551343&0.2453514&0.3053590&0.3485556&0.3813253&0.4071348&0.4280443&0.4453617\\ \hline
\end{tabular}
\end{center}}
{\scriptsize
\begin{center}
\begin{tabular}{||c|c|c|c|c|c|c|c|c||}
\hline
\multicolumn{9}{||c||}{$\lambda=1$}\\
\hline
$\eta_0 \downarrow/ \theta\beta v^4\rightarrow$&0.25&0.50&0.75&1.00&1.25&1.50&1.75&2.00\\
\hline
 0.1&0.1090283&0.2173627&0.3166714&0.4021973&0.4731451&0.5311410&0.5785618&0.6176438\\ \hline
 0.2&0.1109311&0.2183089&0.3147740&0.3967814&0.4643727&0.5195071&0.5645938&0.6017931\\ \hline
 0.3&0.1128118&0.2178051&0.3093577&0.3857451&0.4481253&0.4988503&0.5403345&0.5746132\\ \hline
 0.4&0.1131022&0.2134202&0.2978589&0.3668159&0.4225433&0.4676949&0.5046197&0.5351800\\ \hline
 0.5&0.1101025&0.2026902&0.2777854&0.3378061&0.3858145&0.4245707&0.4562607&0.4825283\\ \hline
 0.6&0.1021902&0.1834874&0.2471040&0.2969560&0.3364634&0.3682520&0.3942401&0.4158104\\ \hline
\end{tabular}
\end{center}}
{\scriptsize
\begin{center}
\begin{tabular}{||c|c|c|c|c|c|c|c|c||}
\hline
\multicolumn{9}{||c||}{$\lambda=0.2,\hspace{1cm}\beta v^2=0.25$}\\
\hline
$\eta_0\downarrow / \theta v^2\rightarrow$&0.25&0.50&0.75&1.00&1.25&1.50&1.75&2.00\\
\hline
 0.1&0.2200395&0.3539169&0.4451356&0.5118180&0.5629637&0.6035890&0.6367285&0.6643341\\ \hline
 0.2&0.2125212&0.3421194&0.4305462&0.4952475&0.5449058&0.5843679&0.6165695&0.6434008\\ \hline
 0.3&0.2000783&0.3225674&0.4063440&0.4677398&0.5149130&0.5524300&0.5830621&0.6085969\\ \hline
 0.4&0.1828486&0.2954343&0.3727072&0.4294674&0.4731484&0.5079279&0.5363488&0.5600553\\ \hline
 0.5&0.1610389&0.2609806&0.3299035&0.3806889&0.4198565&0.4510904&0.4766430&0.4979747\\ \hline
 0.6&0.1349428&0.2195764&0.2783132&0.3217720&0.3553841&0.3822418&0.4042464&0.4226362\\ \hline
\end{tabular}
\end{center}}
{\scriptsize
\begin{center}
\begin{tabular}{||c|c|c|c|c|c|c|c|c||}
\hline
\multicolumn{9}{||c||}{$\lambda=0.2,\hspace{1cm}\beta v^2=1.75$}\\
\hline
$\eta_0\downarrow / \theta v^2\rightarrow$&0.25&0.50&0.75&1.00&1.25&1.50&1.75&2.00\\
\hline
 0.1&0.2448837&0.3915553&0.4890137&0.5584838&0.6105600&0.6510941&0.6835752&0.7102112\\ \hline
 0.2&0.2374215&0.3795154&0.4739330&0.5412533&0.5917339&0.6310380&0.6625417&0.6883821\\ \hline
 0.3&0.2249399&0.3594183&0.4487860&0.5125366&0.5603667&0.5976261&0.6275043&0.6520206\\ \hline
 0.4&0.2073859&0.3312331&0.4135669&0.4723459&0.5164819&0.5508890&0.5784973&0.6011628\\ \hline
 0.5&0.1847135&0.2949476&0.3682952&0.4207213&0.4601320&0.4908856&0.5155825&0.5358716\\ \hline
 0.6&0.1569152&0.2506022&0.3130459&0.3577574&0.3914206&0.4177226&0.4388662&0.4562506\\ \hline
\end{tabular}
\end{center}}
{\scriptsize
\begin{center}
\begin{tabular}{||c|c|c|c|c|c|c|c|c||}
\hline
\multicolumn{9}{||c||}{$\lambda=0.5,\hspace{1cm}\beta v^2=0.25$}\\
\hline
$\eta_0\downarrow / \theta v^2\rightarrow$&0.25&0.50&0.75&1.00&1.25&1.50&1.75&2.00\\
\hline
 0.1&0.1602409&0.2737020&0.3585048&0.4244411&0.4772755&0.5206269&0.5568846&0.5876912\\ \hline
 0.2&0.1550665&0.2649169&0.3470589&0.4109533&0.4621707&0.5042094&0.5393793&0.5692690\\ \hline
 0.3&0.1464564&0.2503017&0.3280186&0.3885166&0.4370436&0.4768968&0.5102548&0.5386168\\ \hline
 0.4&0.1344382&0.2299062&0.3014492&0.3572060&0.4019748&0.4387729&0.4695963&0.4958196\\ \hline
 0.5&0.1190664&0.2038210&0.2674641&0.3171484&0.3570982&0.3899747&0.4175409&0.4410131\\ \hline
 0.6&0.1004407&0.1722034&0.2262524&0.2685494&0.3026267&0.3307163&0.3543005&0.3744046\\ \hline
\end{tabular}
\end{center}}
{\scriptsize
\begin{center}
\begin{tabular}{||c|c|c|c|c|c|c|c|c||}
\hline
\multicolumn{9}{||c||}{$\lambda=0.5,\hspace{1cm}\beta v^2=1.75$}\\
\hline
$\eta_0\downarrow/ \theta v^2\rightarrow$&0.25&0.50&0.75&1.00&1.25&1.50&1.75&2.00\\
\hline
 0.1&0.2283353&0.3828428&0.4892865&0.5653020&0.6216879&0.6649564&0.6991253&0.7267608\\ \hline
 0.2&0.2233610&0.3731310&0.4759154&0.5492336&0.6036180&0.6453673&0.6783533&0.7050458\\ \hline
 0.3&0.2146406&0.3564993&0.4532769&0.5221908&0.5733087&0.6125753&0.6436251&0.6687711\\ \hline
 0.4&0.2015959&0.3323599&0.4209074&0.4838293&0.5305060&0.5663915&0.5947978&0.6178272\\ \hline
 0.5&0.1835371&0.3000318&0.3782747&0.4337543&0.4749187&0.5065990&0.5317079&0.5520893\\ \hline
 0.6&0.1597608&0.2588504&0.3248680&0.3715902&0.4062713&0.4329942&0.4542035&0.4714428\\ \hline
\end{tabular}
\end{center}}
{\scriptsize
\begin{center}
\begin{tabular}{||c|c|c|c|c|c|c|c|c||}
\hline
\multicolumn{9}{||c||}{$\lambda=0.8,\hspace{1cm}\beta v^2=0.25$}\\
\hline
$\eta_0\downarrow / \theta v^2\rightarrow$&0.25&0.50&0.75&1.00&1.25&1.50&1.75&2.00\\
\hline
 0.1&0.0867972&0.1612771&0.2255448&0.2813263&0.3300317&0.3728119&0.4106060&0.4441825\\ \hline
 0.2&0.0847113&0.1572219&0.2196691&0.2737899&0.3209924&0.3624182&0.3989940&0.4314747\\ \hline
 0.3&0.0811168&0.1502752&0.2096520&0.2609908&0.3056882&0.3448651&0.3794237&0.4100937\\ \hline
 0.4&0.0758505&0.1401799&0.1951888&0.2426072&0.2837994&0.3198458&0.3516071&0.3797730\\ \hline
 0.5&0.0687112&0.1266240&0.1759151&0.2182588&0.2549512&0.2870036&0.3152118&0.3402070\\ \hline
 0.6&0.0594862&0.1092859&0.1514638&0.1875696&0.2187798&0.2459975&0.2699244&0.2911117\\ \hline
\end{tabular}
\end{center}}
{\scriptsize
\begin{center}
\begin{tabular}{||c|c|c|c|c|c|c|c|c||}
\hline
\multicolumn{9}{||c||}{$\lambda=0.8,\hspace{1cm}\beta v^2=1.75$}\\
\hline
$\eta_0\downarrow/ \theta v^2\rightarrow$&0.25&0.50&0.75&1.00&1.25&1.50&1.75&2.00\\
\hline
 0.1&0.2077608&0.3713083&0.4891521&0.5729292&0.6338726&0.6796519&0.7151113&0.7433168\\ \hline
 0.2&0.2061466&0.3647761&0.4778972&0.5581057&0.6164778&0.6603812&0.6944332&0.7215522\\ \hline
 0.3&0.2023295&0.3526599&0.4582084&0.5327548&0.5870391&0.6279448&0.6597363&0.6851018\\ \hline
 0.4&0.1948819&0.3334512&0.4289408&0.4960701&0.5449875&0.5819318&0.6107156&0.6337338\\ \hline
 0.5&0.1822157&0.3055486&0.3888778&0.4471780&0.4896931&0.5218790&0.5470224&0.5671795\\ \hline
 0.6&0.1628170&0.2675205&0.3369334&0.3852825&0.4205688&0.4473456&0.4683193&0.4851761\\ \hline
\end{tabular}
\end{center}}
\subsection{Comparison with the semilocal commmutative vortices}
Let us now compare with the commutative semilocal model. With the radial ansatz
\beqrn
\phi^+&=&v g(r) e^{i n \varphi}\\
\phi^0&=&v h(r) e^{i l \varphi}\\
A_\theta&=&n-\alpha(r)
\eeqrn
the commutative Bogomolny equations  are
\beqr
\frac{1}{r}\frac{d\alpha}{dr}&=&\left[1-\lambda+\lambda\beta v^2 \left(g^2+h^2\right)\right](g^2+h^2-1)\label{eq:da}\\
\frac{dg}{dr}&=&\frac{\alpha g}{r} \label{eq:dg}\\
\frac{dh}{dr}&=&\frac{\alpha-n+l}{r} h \label{eq:dh} 
\eeqr 
and the boundary conditions are
\beqrn
&g(0)=0&\hspace{3cm}h(0)=h_0\delta_{l,0}\hspace{2.5cm}\alpha(0)=n\\
&g(\infty)=0&\hspace{3cm}h(\infty)=0\hspace{3cm}\alpha(\infty)=0 .
\eeqrn
Let us concentrate in the case $n=1$, $l=0$. From (\ref{eq:dg}) and (\ref{eq:dh}), it follows that $h(r)=\frac{\rho}{r} g(r)$ with $\rho=\frac{h_0}{g^\prime(0)}$. Using this in (\ref{eq:da}) one can see that the solution has the form
\beqrn
g(r)&\simeq&g_0 r\\
\alpha(r)&\simeq&1+\frac{1}{2}\left(1+\lambda\beta v^2 |h_0|^2-\lambda\right)\left(|h_0|^2-1\right) r^2
\eeqrn
when $r\simeq 0$. The integration of the equations by the Runge-Kutta method shows that the values of $g_0^2$ compatible with the boundary conditions at infinity are the following: 
{\scriptsize
\begin{center}
\begin{tabular}{||c|c|c|c|c|c|c||}
\hline
\multicolumn{7}{||c||}{$g0^2$ for commutative semilocal vortices }\\
\hline
&\multicolumn{6}{|c||}{$\eta0$}\\
\cline{2-7}
Parameter values $\downarrow$&0.1&0.2&0.3&0.4&0.5&0.6\\
\hline
$\lambda=0$ &0.7191&0.6928&0.6494&0.5898&0.5151&0.4269\\ \hline
$\lambda=0.2,\beta v^2=0.25$ &0.5863&0.5655&0.5311&0.4836&0.4238&0.3527\\ \hline
$\lambda=0.2,\beta v^2=1.75$ &0.6522&0.6328&0.6002&0.5541&0.4942&0.4201\\ \hline
$\lambda=0.8,\beta v^2=0.25$ &0.1875&0.1833&0.1759&0.1650&0.1500&0.1304\\ \hline
$\lambda=0.8,\beta v^2=1.75$ &0.4444&0.4478&0.4501&0.4469&0.4327&0.4015\\ \hline
$\lambda=1,\beta v^2=0.25$ &0.0531&0.0548&0.0570&0.0587&0.0590&0.0566\\ \hline
$\lambda=1,\beta v^2=1.75$ &0.3717&0.3838&0.3991&0.4111&0.4126&0.3958\\ \hline
\end{tabular}
\end{center}}
\noindent We have checked that these are precisely the values of $\frac{f_0^2}{2\theta v^2}$ obtained in the noncommutative model when we take the limit $\theta\rightarrow 0$, as it should be.
\subsection{Field profiles of the semilocal noncommutative vortices}
Finally, by applying the Weyl transform we can find the profiles of the semilocal vortices for different values of the paremeters. The formulas are
\beqrn
\phi^+(x)\ast\bar{\phi}^+(x)&=&v^2\sum_{k=0}^\infty f_k^2 f_{j,j}(x)\\
\phi^0(x)\ast\bar{\phi}^0(x)&=&v^2\sum_{k=0}^\infty \eta_k^2 f_{j,j}(x)\\
\eeqrn
and
\bdm
B(x)=v^2\sum_{k=0}^\infty (1-\lambda+\lambda\beta v^2 (f_k^2+\eta_k^2))(1-f_k^2-\eta_k^2) f_{j,j}(x)\ .
\edm
We illustrate the results for several cases in the following figures, where \mbox{$\phi^+(x)\ast\bar{\phi}^+(x)$},  \mbox{$\phi^0(x)\ast\bar{\phi}^0(x)$} and $B(x)$ are plotted, respectively, in red, blue and green.
\begin{figure}[H]
\centering
\includegraphics[width=7.5cm,height=7cm]{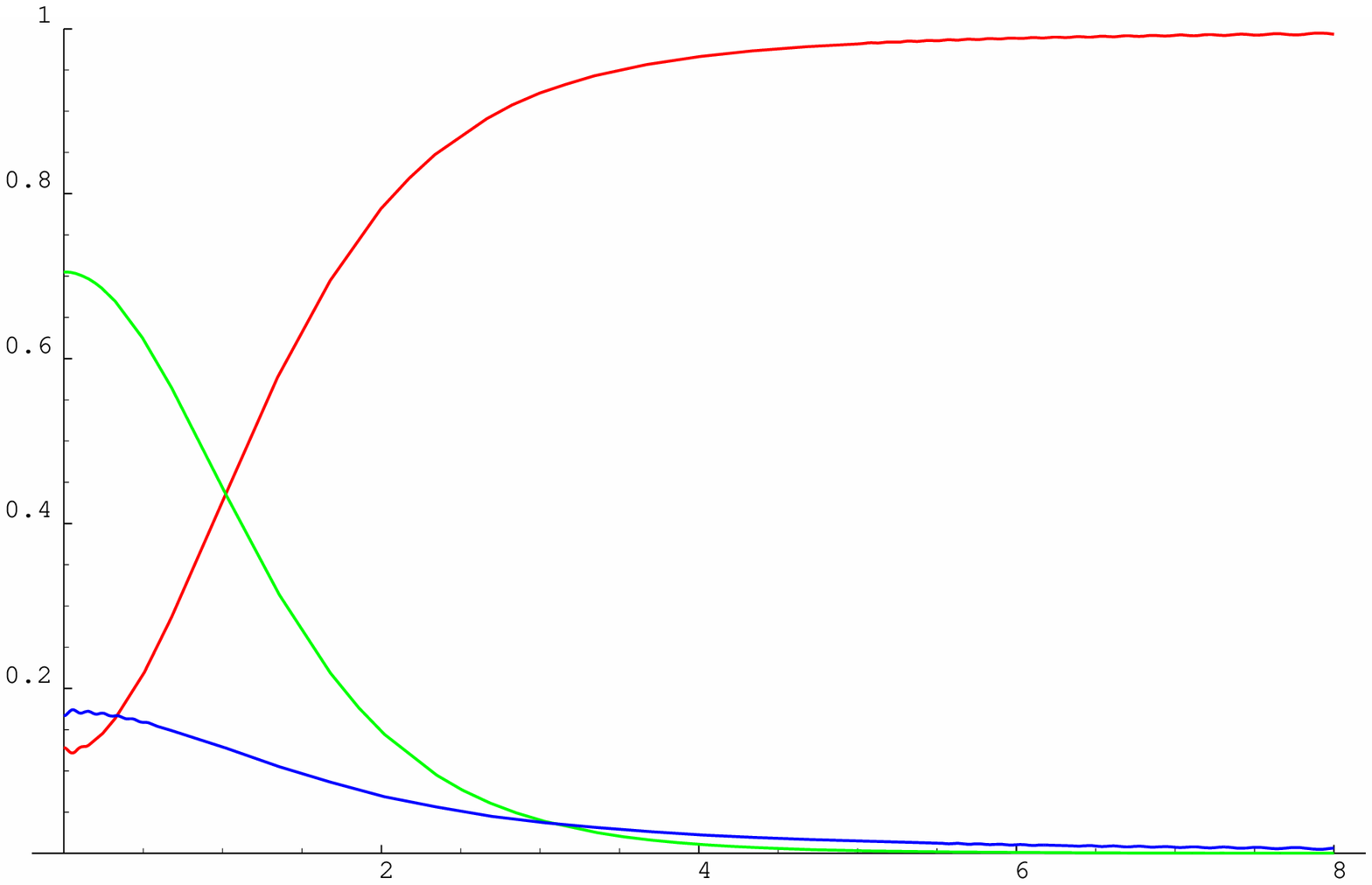}
\includegraphics[width=7.5cm,height=7cm]{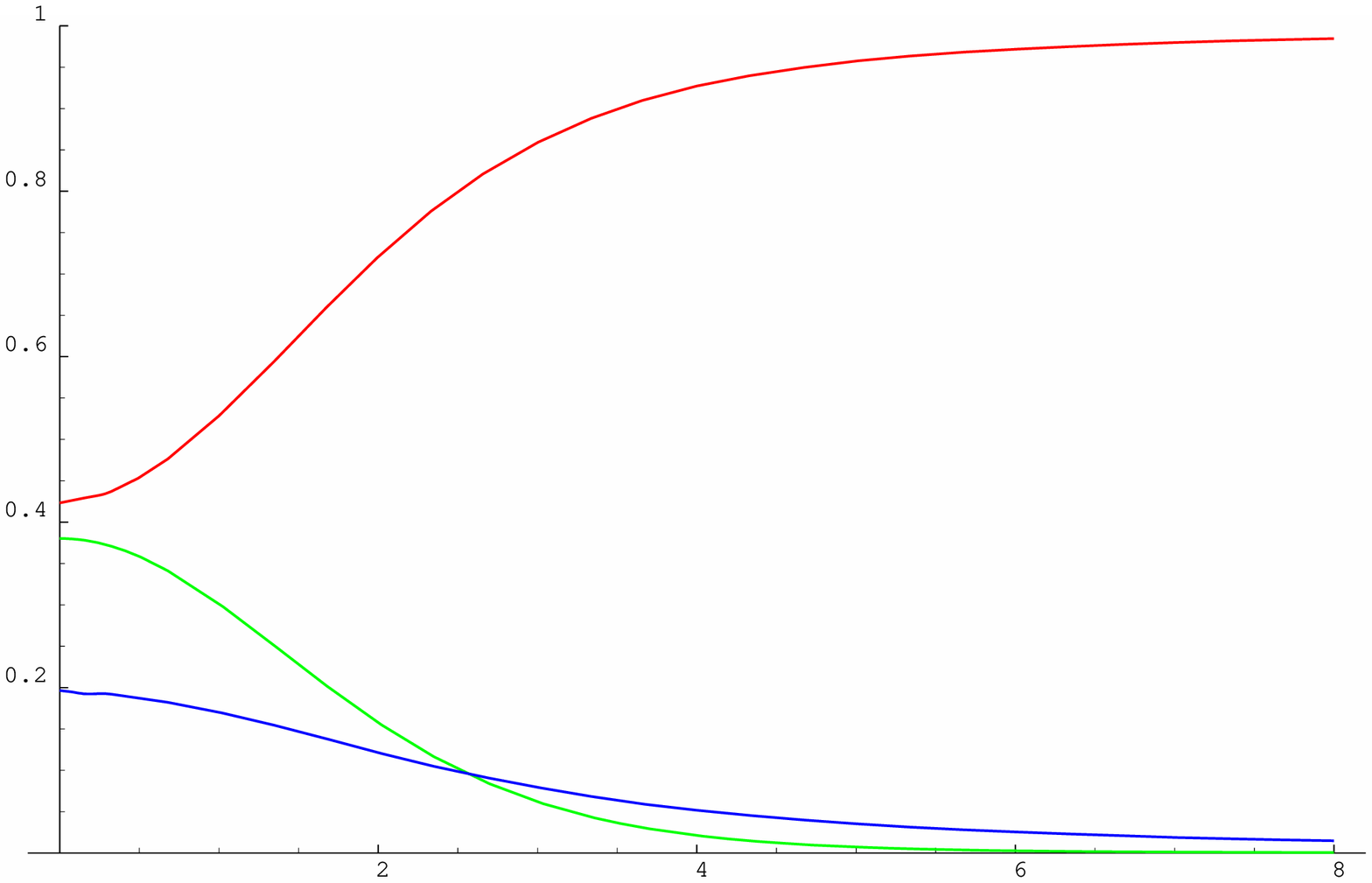}
\caption{The cases $\lambda=0,\eta_0=0.4, \theta v^2=0.25$ (left) and $\lambda=0,\eta_0=0.4, \theta v^2=1.75$ (right).}
\end{figure}
\begin{figure}[H]
\centering
\includegraphics[width=7.5cm,height=7cm]{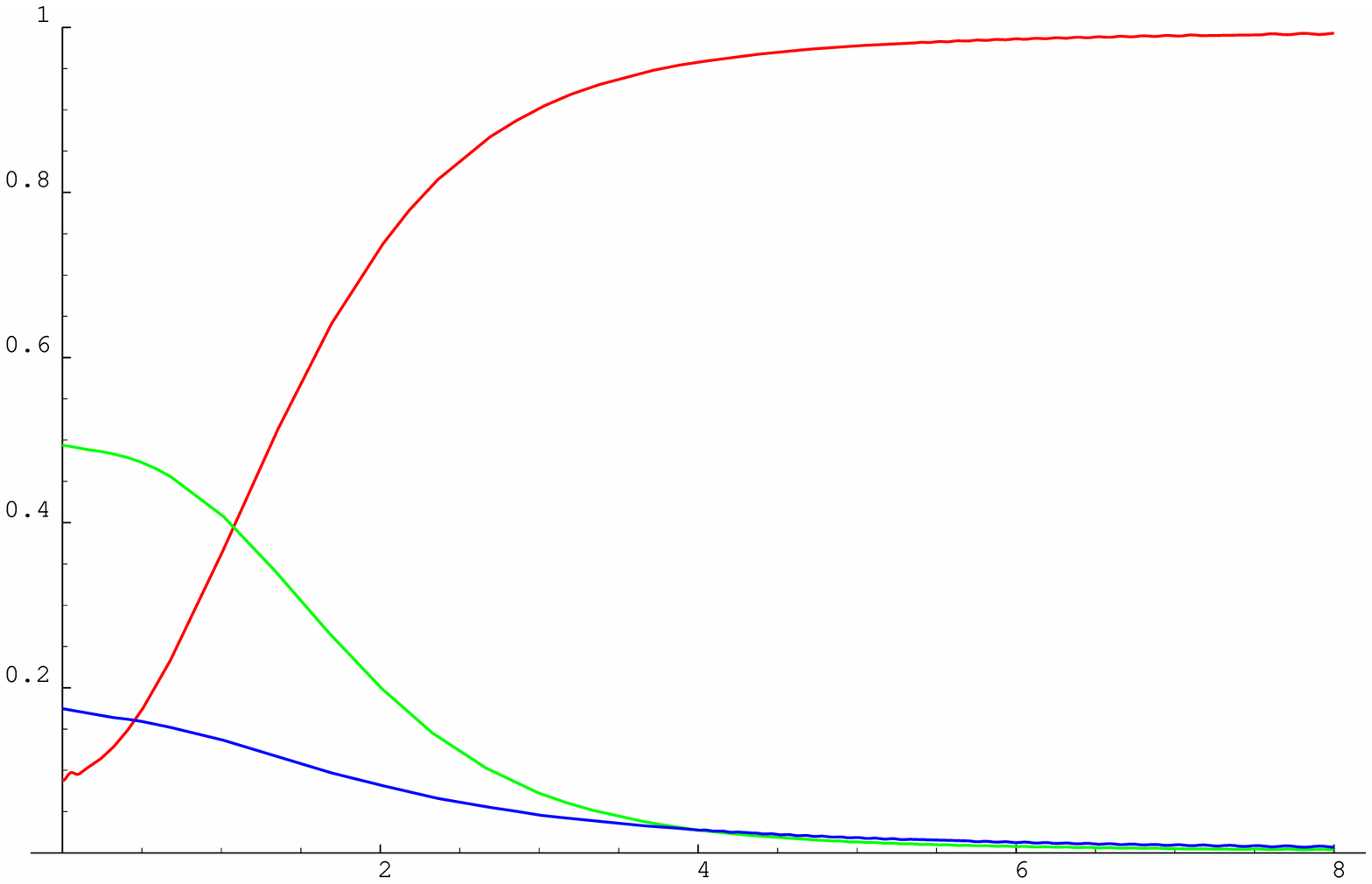}
\includegraphics[width=7.5cm,height=7cm]{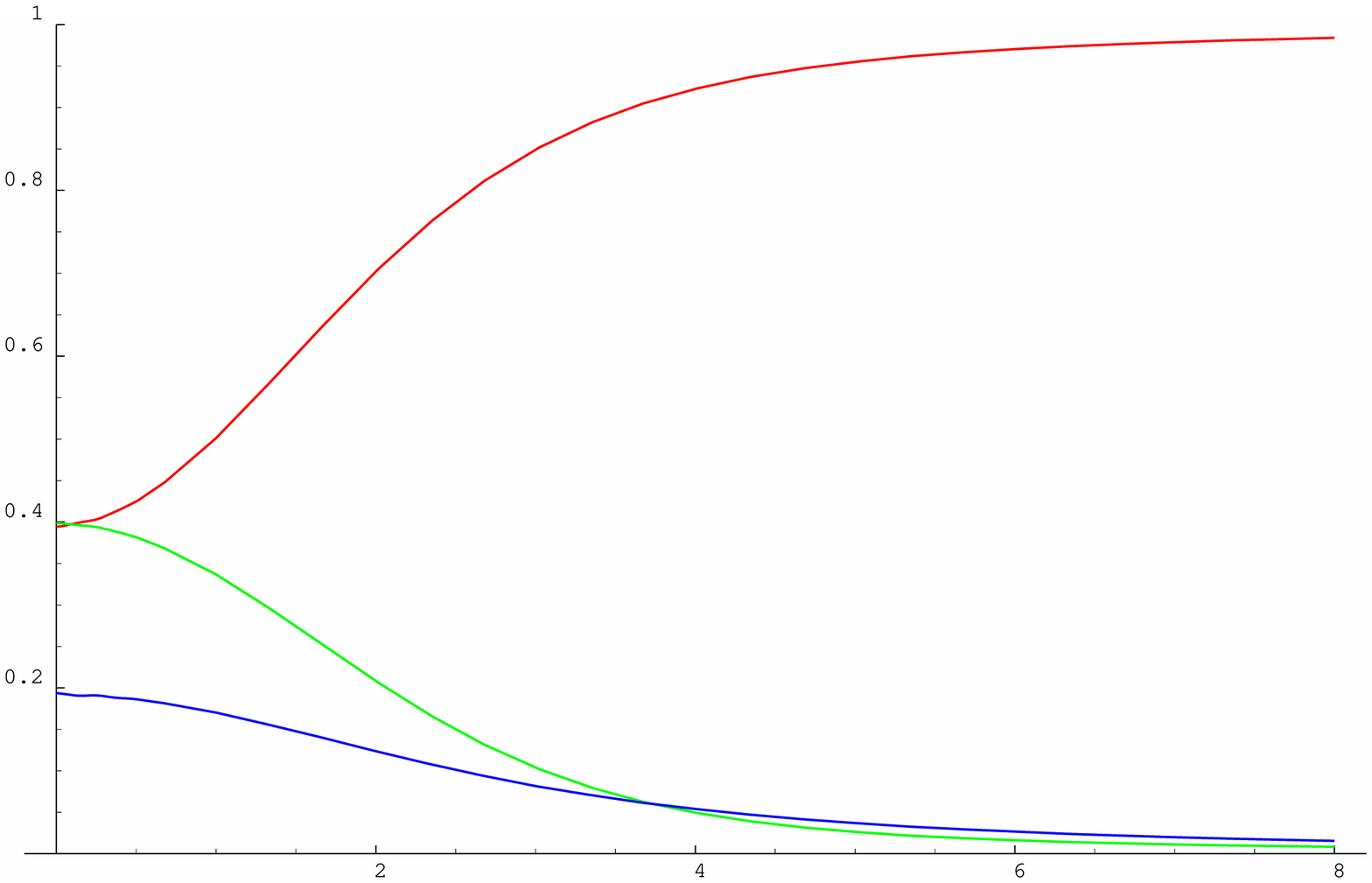}
\caption{The cases $\lambda=0.5,\eta_0=0.4, \theta v^2=0.25,\beta v^2=1$ (left) and $\lambda=0.5,\eta_0=0.4, \theta v^2=1.75,\beta v^2=1$ (right).}
\end{figure}
\begin{figure}[H]
\centering
\includegraphics[width=7.5cm,height=7cm]{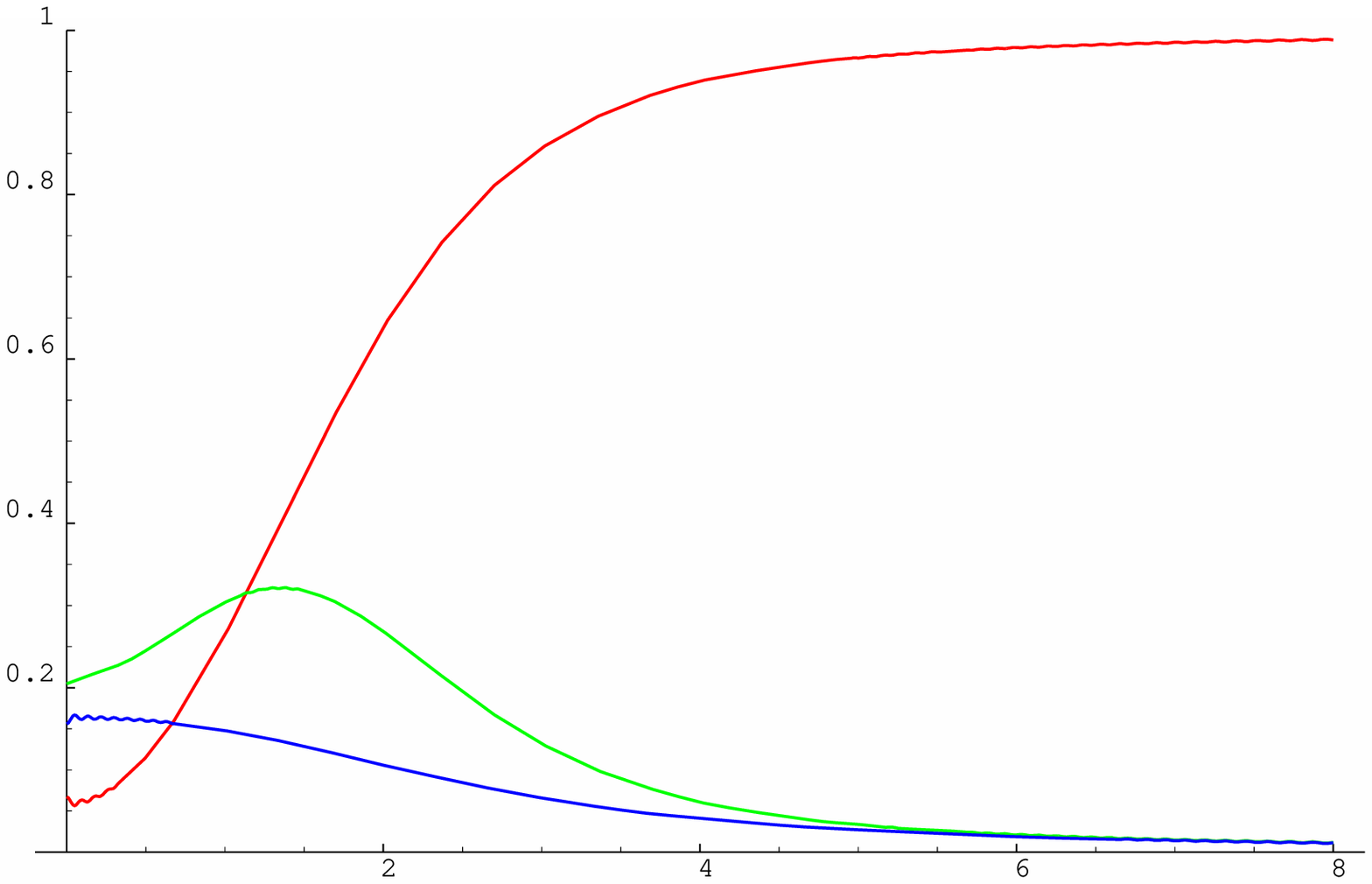}
\includegraphics[width=7.5cm,height=7cm]{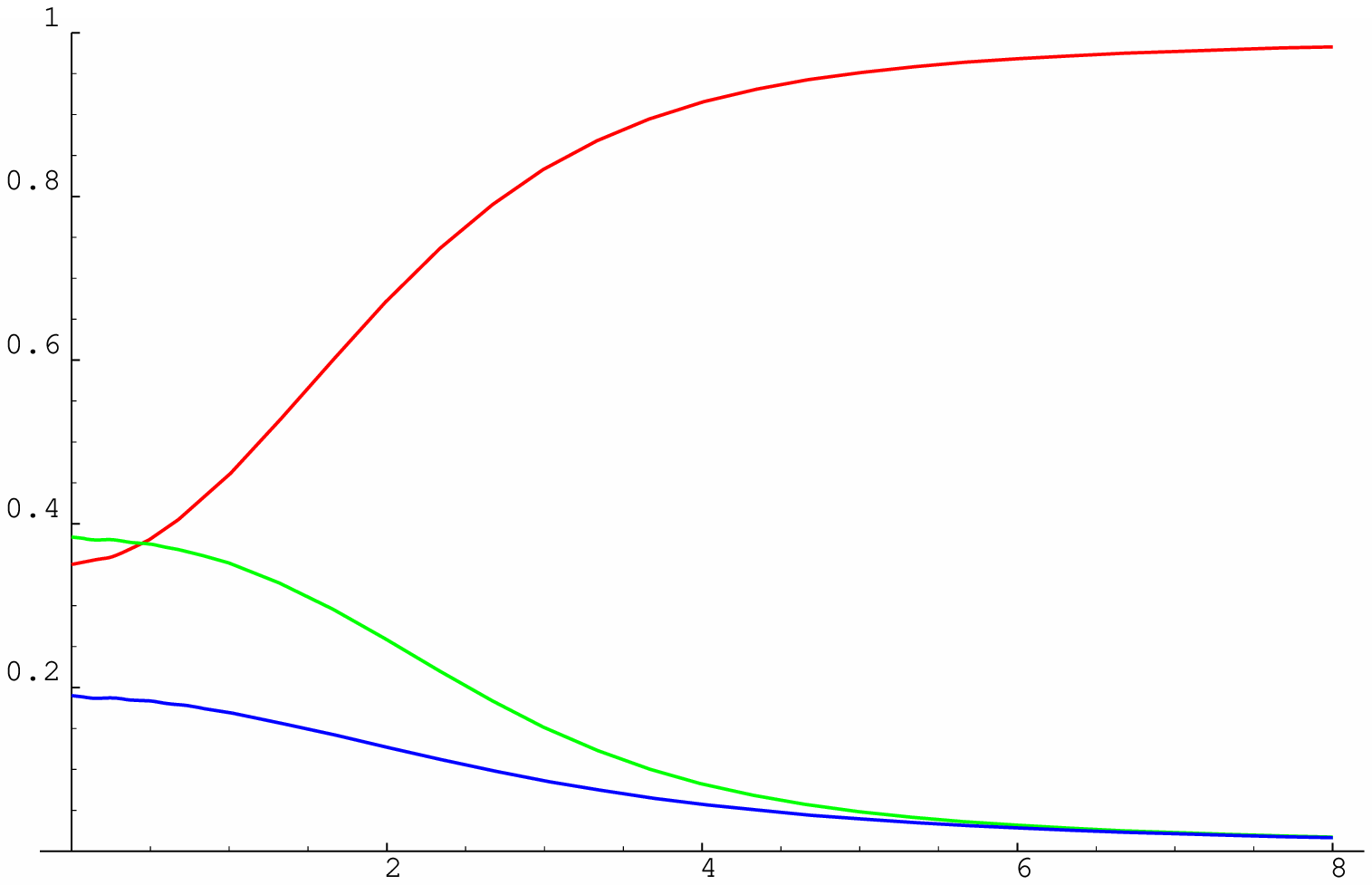}
\caption{The cases $\lambda=1,\eta_0=0.4, \theta v^2=0.25,\beta v^2=1$ (left) and $\lambda=1,\eta_0=0.4, \theta v^2=1.75,\beta v^2=1$ (right).}
\end{figure}
\noindent In these examples we have chosen an intermediate value for $\eta_0$ and three values for the interpolating parameter, corresponding to semilocal Nielsen-Olesen vortices, semilocal Chern-Simons vortices, and a third case just in the middle of the range of $\lambda$. We have fixed the $\beta$ parameter to a value $\beta v^2=1$ and present, for each value of $\lambda$, solutions for both small ($\theta v^2=0.25$) and large ($\theta v^2=1.75$) noncommutativities. Some features that we can appreciate looking at the figures are as follows:
\begin{itemize}
\item In the case of small $\theta v^2$ the profile of the magnetic field has a maximum at the center of the Nielsen-Olesen vortex, but it is ring-shaped for the Chern-Simons case; for the intermediate $\lambda=0.5$ solution the maximum is still there, although more flat. 
\item This pattern changes when the noncommutativity is large. In this case, the magnetic field profiles for $\lambda=0$ and $\lambda=0.5$ are nearly the same and, although with a slightly more flat maximum, the magnetic field remains concentrated at the core of the vortex also for $\lambda=1$. 
\item  The magnitude of the upper component of the scalar field is minimum at the origin. The value of \mbox{$\phi^+(0)\ast\bar{\phi}^+(0)$} decreases with $\lambda$, both for small and large $\theta$. 
\item For the three values of $\lambda$, \mbox{$\phi^+(0)\ast\bar{\phi}^+(0)$} is higher for larger noncommutativity.  Accordingly, the growth with distance of \mbox{$\phi^+(x)\ast\bar{\phi}^+(x)$} is less steep in that case. 
\item The magnitude of the lower component of the scalar field has a maximum at the vortex core.  There is, for the three values of $\lambda$, some increase of the value of \mbox{$\phi^0(0)\ast\bar{\phi}^0(0)$} when $\theta v^2=1.75$ as compared with $\theta v^2=0.25$, but the effect is  small. In all cases, \mbox{$\phi^0(x)\ast\bar{\phi}^0(x)$} converges quite slowly to its asymptotic value, and the higher the noncommutativity, the slower the convergence.
\end{itemize}
\section{Conclusions and outlook}
In this paper we have studied the standard and semilocal noncommutative generalized AHM with dielectric function, showing that they admit a Bogomolny splitting and have, therefore, stable vorticial solutions whose energy is proportional to the magnetic flux. By changing the dielectric function it is possible to model the vortices in a variety of shapes. For the $U(1)$ model, we have focused in the case of unit vorticity and provided a number of solutions with different values of the non-dimensional parameters $\theta v^2$ and $\beta v^2$, finding vorticial profiles which interpolate between those of the noncommutative Nielsen-Olesen and Chern-Simons-Higgs cases. We have checked numerically that, for the case of $\theta\rightarrow 0$, regular solutions exist which converge to the vortices of the commutative model. The noncommutative $SU(2)\times U(1)$ semilocal model with dielectric function has also been investigated along the same lines, and their self-duality equations have been solved numerically for a variety of values of the above mentioned parameters and also of the coefficient $\eta_0$ which measures the degree of departure between standard and semilocal vortices.

Finally, let us  make a couple of comments on some possible directions to extend this work in future research. Here we have concentrated in the case of a single vortex but, as it is well known~\cite{jatau}, the commutative self-duality equations admit multivortex solutions spanning a moduli space which has dimension $2n$ in the topological sector of winding number~$n$~\cite{wein}. It would be interesting to elaborate on the generalization of this result to the noncommutative cases with dielectric function that we have been studying. For the $U(1)$ case, for instance, if we shift the scalar and gauge fields of a vortex under the condition that the self-duality equations continue to be satisfied to linear order in the deformations $\delta\phi, \delta A_k$, we arrive to a equation of the form
\bdm
{\cal D} \Psi=0
\edm
where 
\bdm
{\cal D}=\left(\begin{array}{cccc}
D_1+iD_2&0&-i R_\phi &R_\phi\\
0&\bar{D}_1+i\bar{D}_2&i L_{\bar\phi} &L_{\bar\phi}\\
\nabla_2&-\nabla_1&U_{R_{\bar{\phi}}}&U_{L_\phi}\\
\nabla_1&\nabla_2&i R_{\bar\phi} &-i L_\phi
\end{array}\right)
\hspace{0.7cm}{\rm and}\hspace{0.7cm}
\Psi=\left(\begin{array}{c}
\delta\phi\\
\delta\bar\phi\\
\delta A_1\\
\delta A_2
\end{array}\right).
\edm
The first three rows in ${\cal D}$ come form the linearization of (\ref{eq:dual1})-(\ref{eq:dual2}), whereas the fourth one is a background gauge condition suitable to remove the spuriuos deformations which amount only to a change of gauge. The elements of ${\cal D}$ are operators whose action on the deformations is as follows:
\beqrn
D_k \delta X&=&\partial_k \delta X- i A_k\ast \delta X\\
\bar{D}_k \delta X&=&\partial_k \delta X+ i \delta X\ast A_k\\
\nabla_k \delta X&=&\partial_k \delta X-i A_k\ast \delta X+ i \delta X\ast A_k\\
R_Y \delta X&=& \delta X\ast Y\hspace{0.8cm}L_Y \delta X=Y\ast \delta X\\
U_{K_Y}\delta X&=&\lambda\beta (K_Y \delta X)\ast(v^2-\phi\ast\bar{\phi})-[(1-\lambda)+\lambda\beta\phi\ast\bar{\phi}]\ast (K_Y \delta X) .
\eeqrn
In the commutative case, the dimension of the vortex moduli space ${\cal M}$ is given by the index of ${\cal D}$ \cite{wein}
\bdm
\dim {\cal M}={\rm ind}{\cal D}=\dim {\rm ker}{\cal D}-\dim {\rm ker}{\cal D}^\dagger 
\edm
and, given that this is a topological quantity, and the noncommutative self-duality equations are continuous deformations in the $\theta$ parameter of the commutative ones, we expect on general grounds that the result valid for commutative vortices is still  valid for any $\theta$. Nevertheless, all the details of the computation, such as to establish a vanishing theorem for the kernel of ${\cal D}^\dagger$ or to evaluate the heat-kernel traces of the superpartner Laplacians ${\cal D}^\dagger{\cal D}$ and ${\cal D}{\cal D}^\dagger$, seem to be quite intricate for objects involving the Groenewold-Moyal product, see \cite{vass2,vass3}. In particular, the coefficients of the asymptotic expansions of these heat-kernel traces split into three terms 
\bdm
a_n({\cal O})=a_n^L({\cal O})+a_n^R({\cal O})+a_n^{\rm mix}({\cal O})\hspace{1cm}{\rm for}\hspace{1cm}{\cal O}={\cal D}^\dagger{\cal D}\hspace{0.5cm}{\rm or}\hspace{0.5cm} {\cal D}{\cal D}^\dagger ,
\edm
where $a_n^L({\cal O})$ involves only the fields entering in ${\cal O}$ as left Moyal multipliers, $a_n^R({\cal O})$~includes only right Moyal multipliers, and  $a_n^{\rm mix}({\cal O})$ is given by a combination of fields of both types. Furthermore, this last term is divergent as $\theta^{-1}$ when the commutativity of the plane is restored. Thus, an issue to be clarified is if the good behavior in the limit $\theta\rightarrow 0$ of the solutions reported here is enough to ensure that the coefficients $a_n^{\rm mix}$ coming from the deformation operator ${\cal D}$ effectively vanish. On the other hand, the heat-kernel coefficients are interesting also from the point of view of computing the quantum corrections to the semiclassical energy of vortices. In fact, the main part of this correction comes from the trace of ${\cal D}^\dagger{\cal D}$ once a convenient regularization scheme, based for instance in zeta-function methods~\cite{pos9}, is stipulated. For the commutative $U(1)$ and semilocal vortices, the computation of the leading $a_n({\cal D}^\dagger{\cal D})$ needeed for the quantum corrections has been performed in \cite{zeta}. We think that it would be a worthwhile project to study the precise way in which the methods described there can be generalized in order to be applied to the case of  noncommutative solutions.

\end{document}